\documentclass[preprint2, trackchanges, twocolumn]{aastex701}

\usepackage{mathtools}
\usepackage{todonotes}
\usepackage{xcolor}
\usepackage{booktabs}
\usepackage{verbatim}
\usepackage{color, soul}

\def\code#1{\texttt{#1}}

\newcommand{\eg}[1]{(e.g. \citealt{#1})}
\defcitealias{EHTM87Paper5}{EHTC~M87*~V}
\defcitealias{EHTSgrAPaper5}{EHTC~Sgr~A*~V}
\defcitealias{EHTM87Paper7}{EHTC~M87*~VII}

\begin{document}
\title{\texttt{GPUmonty}: A GPU-accelerated relativistic Monte Carlo radiative transfer code}

\author[orcid=0000-0002-9318-9329]{Pedro Naethe Motta}
\email[show]{pedronaethemotta@usp.br}
\affiliation{Instituto de Astronomia, Geof\'{\i}sica e Ci\^encias Atmosf\'ericas, Universidade de S\~ao Paulo, S\~ao Paulo, SP 05508-090, Brazil.}
\correspondingauthor{Pedro Naethe Motta}

\author[orcid=0000-0003-3956-0331]{Rodrigo Nemmen}
\affiliation{Instituto de Astronomia, Geof\'{\i}sica e Ci\^encias Atmosf\'ericas, Universidade de S\~ao Paulo, S\~ao Paulo, SP 05508-090, Brazil.}
\email[]{rodrigo.nemmen@iag.usp.br}

\author[orcid=0000-0002-2514-5965]{Abhishek V. Joshi}
\affiliation{Department of Physics, University of Illinois, 1110 West Green Street, Urbana, IL 61801, USA}
\email[]{avjoshi2@illinois.edu}

\begin{abstract}
We introduce \code{GPUmonty}, a CUDA/C-based Monte Carlo radiative transfer code accelerated using graphics processing units (GPUs). \code{GPUmonty} derives from the CPU-based code \code{grmonty} and offloads the most computationally expensive stages of the calculation---superphoton generation, sampling, tracking, and scattering---to the GPU. Whereas \code{grmonty} handles photons sequentially, \code{GPUmonty} processes large numbers of superphotons concurrently, leveraging the single-instruction, multiple-thread (SIMT) execution model of modern GPUs. Benchmarks demonstrate a speedup of about $12\times$ relative to the original CPU implementation on a single GPU, with runtime limited primarily by register pressure rather than compute or memory bandwidth saturation. We validate the implementation through analytic tests for a optically thin synchrotron sphere, as well as comparisons with \code{igrmonty} for scattering synchrotron sphere and GRMHD simulation data. Relative errors remain below a percent level and convergence is consistent with the expected $N_{\rm s}^{-1/2}$ Monte Carlo scaling. By significantly reducing computational costs, GPUmonty enables the extensive parameter space surveys and faster spectra modeling required to interpret horizon-scale observations of supermassive black holes. \code{GPUmonty} is publicly available under the GNU General Public License.
\end{abstract}

\section{Introduction}

To connect the multitude of multiwavelength observations of accreting black holes with their physics and perform parameter estimation, we need to model the electromagnetic spectrum that emerges from the hot gas flowing in a curved spacetime. There are a number of techniques for doing so. 
The two main techniques, emitter-to-observer and observer-to-emitter, involve solving the radiative transfer equation and postprocessing snapshots of the fluid and magnetic field quantities generated from general relativistic magnetohydrodynamic (GRMHD) simulations~~\mbox{\eg{Wong:2022rqr}}

The workhorse method for generating images and spectra in the radio and sub-millimeter regime relevant for VLBI imaging of Sgr A* or M87* (\citealt{EHTC2019, EHTC2022}) is backward ray tracing. Here, instead of tracking every photon---most of which never reach the observer---we define a virtual ``camera'' far from the black hole. The trajectory of the rays are calculated backwards from each pixel of the camera toward the black hole horizon. Examples of ray tracing implementations of this observer-to-emitter include \code{ipole} \citep{Monika_2018}, \code{BHOSS} \citep{Younsi_2012}, \code{Jipole}~\citep{Naethe_Motta_2025} and \code{GRay} \citep{Chan2013}. Backward ray tracing struggles when there is significant scattering---relevant when modeling hard X-ray emission from accretion flows or gamma-ray production in relativistic jets.

In such cases, the standard technique employs Monte Carlo methods. Here, we generate  probabilistically superphotons---packets representing many photons---throughout the flow based on the local emissivity. The gold standard implementation of this method is \code{grmonty} \citep{Dolence_2009}, on which this work is based. Other examples include Pandurata \citep{Schnittman2013}, the multi-CPU \code{grmonty} extension \code{$\kappa$monty} \citep{Davelaar2023}, and \code{RAIKOU} \citep{Kawashima2023}, which uses the observer-to-emitter formalism to generate images and the reverse procedure to generate spectra

While recent efforts have focused on accelerating synthetic image generation in the context of Event Horizon Telescope observations \citep{Palumbo_2022, Tiede_2022, Moscibrodzka_2023, Sharma:2023nbk, Chang_2024, Yfantis_2024_bipole, Keeble_2025, Naethe_Motta_2025}, multiwavelength spectral calculations---central to constraining models across the electromagnetic spectrum---have received comparatively little attention. Forward-modeling libraries now contain millions of snapshots (60,000 in \citetalias{EHTM87Paper5}; 5.5 million in \citetalias{EHTSgrAPaper5}), yet Monte Carlo radiative transfer remains computationally prohibitive for more detailed analyses: spectra were computed for only 20\% of the M87* images produced in \citetalias{EHTM87Paper5}, owing to the computational expense associated with sampling the large parameter space spanning black hole spin, electron heating prescriptions, and emission models, for example. As a result, Bayesian Monte Carlo Markov chain posterior sampling---now routine for image-based forward modeling---remains impractical for spectral calculations with existing codes. These demands will only intensify with more stringent probes of black hole environment with next-generation instruments such as the ngEHT \citep{Johnson:2023ynn} and space-VLBI missions \citep{Johnson:2024ttr} with their significantly improved angular resolution, dynamic range and cadence, further increasing library sizes and computational costs.

To harness the massively parallel computational power of GPUs for multiwavelength spectral calculations and address the computational bottlenecks previously described, we present \code{GPUmonty}: a relativistic radiative transfer code developed in C/CUDA based on the framework of the Monte Carlo code \code{grmonty}. By exploiting the modern hardware of NVIDIA GPUs, \code{GPUmonty} achieves over $12\times$ speedup compared to \code{grmonty}.

This paper is organized as follows: In Section~\ref{sec:equations}, we provide the governing equations and numerical methods used in \code{GPUmonty}. Section~\ref{sec:gpu_algorithms} describes the implementation of the GPU-accelerated algorithm and the main changes performed when porting \code{grmonty}. In Section~\ref{sec:tests}, we validate the reliability of \code{GPUmonty} by performing three different tests: a uniform optically thin synchrotron sphere, a scattering synchrotron sphere test, and comparing spectra generated with \code{GPUmonty} and \code{grmonty} from a GRMHD simulation. In Section~\ref{sec:gpu_benchmark}, we present a discussion of the code's performance. Finally, in Section~\ref{sec:conclusion} we present the conclusions and future perspectives.

\section{Governing Equations and Numerical Methods}
\label{sec:equations}

\subsection{Creation of superphotons}
We follow the methods used in \code{grmonty} and treat the photon field as a collection of photon packets (so called superphotons), with a weight $w$ representing the number of photons within each packet. The relation between superphotons and physical photons is given by $dN = w dN_{\rm s}$, where $N_{\rm s}$ is the number of superphotons and $N$ is the number of photons. The weight $w(\nu)$ is calculated as a function of the photon frequency $\nu$ in the plasma frame. The probability distribution for the superphotons can be written in terms of the weight $w$, frequency $\nu$ and emissivity $j_\nu$ as 
\begin{equation}
    \frac{1}{\sqrt{-g}} \frac{dN_{\rm s}}{d^3x \, dt \, d\nu \, d\Omega} = \frac{1}{w \sqrt{-g}} \frac{dN}{d^3x \, dt \, d\nu \, d\Omega} = \frac{1}{w} \frac{j_\nu}{h \nu}.
\end{equation}
Here, $g$ is the determinant of the metric, $dt$ is a differential time element, $d\Omega$ the differential solid angle, $d\nu$ the frequency interval, $d^3x$ the volume element, and $h$ is Planck's constant. The spectrum is divided into energy bins, and the corresponding weight for each energy bin can be expressed as
\begin{equation}
    w_\nu(\nu) = \frac{\Delta t \Delta {\rm ln} \nu}{h N_{\rm s}}\int \sqrt{-g} \, d^3x \int j_\nu d\Omega.
    \label{eq:weight_calc}
\end{equation} 

In practice, the computational domain is a grid and superphotons are generated independently in each zone considering the local plasma properties, with weights assigned according to the expected contribution of that zone to the spectrum as given by Eq.~\ref{eq:weight_calc}. In \code{GPUmonty}, the total number of superphotons is controlled by the parameter $N_{\rm s}$, which specifies the target number of superphotons to be generated by it; the actual number produced may differ slightly due to the stochastic sampling procedure but typically remains of the same order of magnitude. 

The superphoton generation (emission) and scattering events all occur in the plasma rest frame and boosting back into the coordinate basis accounts for the relativistic effects.

\subsection{Emissivity}
In \code{GPUmonty}, we currently consider only thermal synchrotron emissivity, appropriate for the hot, optically thin accretion flows relevant to the systems studied here, following \cite{Leung_2011}. The emissivity $j_\nu$ is defined as 
\begin{align}
    j_\nu (\nu, \theta) & =\frac{\sqrt{2} \pi e^2 n_e \nu_s}{3 c K_2(1/\Theta_{\rm e})} (X^{1/2} + 2^{11/12} X^{1/6})^2 \times \notag \\ 
    & \exp(-X^{1/3}),
    \label{eq:emissivity}
\end{align}
where $X \equiv \nu/\nu_s$, 
\begin{equation}
\nu_s = \frac{2}{9} \left(\frac{eB}{2\pi m_e c}\right) \Theta_{\rm e}^2 \sin\theta.  
\end{equation}
Here, $\theta$ is the angle between the magnetic field and the wave vector, $e$ is the electron charge, $n_e$ is the electron density, $c$ is the speed of light, $K_2$ is the modified Bessel function of the second kind, $m_e$ is the electron mass, and
$\Theta_{\rm e} = k_b T_e/(m_e c^2) \gtrapprox 0.5$, where $k_b$ is the Boltzmann constant and $T_e$ is the electron temperature. The plasma properties are defined in the fluid frame. 

In the large temperature limit, for numerical stability we approximate $K_2(1/\Theta_{\rm e})\sim 2 \Theta_{\rm e}^2$ without any significant loss in accuracy.

\subsection{Sampling procedure}
\label{sec:sampling_procedure}

The number of superphotons produced in zone $i$ follows
\begin{equation}
    N_{s,i} = \Delta t  \Delta^3 x \sqrt{-g} \int \int d\nu \, d\Omega \frac{1}{w} \frac{j_\nu}{h \nu}.
\end{equation}
These superphotons are distributed over frequency following the distribution
\begin{equation}
    \frac{dN_{s,i}}{d \, \rm{ln} \nu} = \Delta t \Delta^3 x \sqrt{-g} \frac{1}{hw} \int d\Omega j_\nu,
\end{equation}
and each superphoton is assigned a frequency by rejection sampling. Namely, we start by defining the maximum and minimum frequencies ($\nu_{\rm max}, \nu_{\rm min}$, respectively) before running the simulation. A random number, $\rm{rand}_1$, is drawn from the interval $[0, 1)$. The frequency is then chosen using the formula:
\begin{equation}
    \nu_0 = \exp\left[ \rm{rand}_1 \rm{ln}\left(\frac{\nu_{\rm max}}{\nu_{\rm min}}\right) + \rm{ln}\nu_{\rm min}\right]
\end{equation}

Next, a second random number $\rm{rand}_2$ is drawn and the process repeats until the desired condition is met:
\begin{equation}
\rm{rand}_2 < \frac{\left(\frac{dN_{s,i}}{d\rm{ln}_\nu}\right)_{\nu_0}}{\rm{max}\left(\frac{dN_{s,i}}{d ln \nu}\right)}.
\end{equation}

Afterwards, we sample the photon direction by rejection sampling again. A preliminary value for $\theta$ is selected by sampling $\cos\theta = (2 \rm{rand}_3 - 1)$ from a uniform distribution over the range $[-1, 1)$. Next, a second random variable $\rm{rand}_4$ is drawn from a uniform distribution over $[0, 1)$. The proposed value of $\theta$ is accepted if
\begin{equation}
    {\rm rand}_4 < \frac{j_\nu(\theta)}{j_\nu(\pi/2)}.
\end{equation}

\subsection{The geodesic equation}

The superphotons created will follow null geodesics that can be described by the following set of first-order differential equations
\begin{equation}
\begin{aligned}
    \frac{dx^\mu}{d\lambda} &= k^\mu,\\
    \frac{dk^\mu}{d\lambda} &= - \Gamma^\mu_{\alpha \beta} k^\alpha k^\beta,
\end{aligned}
\label{eq:geodesic-system}
\end{equation}
where $x^\mu$ is the four-position of the superphoton, $k^\mu$ is the four-velocity,  $\lambda$ is the affine parameter,
and $\Gamma^\mu_{\alpha\beta}$ are the Christoffel symbols.

To solve the system of equations (\ref{eq:geodesic-system}), we use the Velocity Verlet algorithm, a second-order numerical integration method \citep{Swope1982}. The algorithm updates the positions using the current velocities and accelerations, then computes the new accelerations and updates the velocities by averaging the old and new accelerations. In the context of Eq.~\ref{eq:geodesic-system}, the algorithm obbeys
\begin{gather}
    x^\mu_{n + 1} = x_n^\mu + k_n^\mu \delta \lambda + \frac{1}{2}\left(\frac{dk^\mu}{d\lambda}\right)_{n},
    \\
    k^\mu_{n + 1, p} = k^\mu_{n} + \left(\frac{dk^\mu}{d\lambda}\right)_{n} \delta \lambda,
    \label{eq:predicted_wavevector}
    \\
    \left(\frac{dk^\mu}{d\lambda}\right)_{n + 1} = - \Gamma^\mu_{\alpha\beta}|_{\rm (at \ x_{n + 1})} k^\alpha_{n+1, p} k^\beta_{n + 1,p},
    \\
     k^\mu_{n + 1} = k^\mu_{n} + \frac{1}{2}\left[ \left(\frac{dk^\mu}{d\lambda}\right)_n + \left(\frac{dk^\mu}{d\lambda}\right)_{n + 1}\right] \delta \lambda,
     \label{eq:calculated_wavevector}
\end{gather}
where $n$ is the step and $\delta \lambda$ is the step size. Here, $k^\mu_{\rm n + 1, p}$ represents the predicted velocity at step $n+1$. This prediction is necessary because the change in the wavevector is dependent on the wavevector itself. \code{GPUmonty} uses the calculated value from equation \ref{eq:calculated_wavevector} to evaluate equation \ref{eq:predicted_wavevector}. This process is repeated until the relative error
\begin{equation}
    \epsilon_{\rm err} = \frac{\lvert k^\mu_{\rm n + 1} - k^\mu_{\rm n + 1,p}\rvert}{k^\mu_{\rm n + 1}}
\end{equation}
is smaller than a configurable parameter of the implementation, which we set to $10^{-3}$.

\subsection{Covariant equation of radiative transfer}

To account for the interaction of matter and photons along the geodesics, we use the covariant form of the unpolarized radiative transfer equation \citep{Mihalas_1984, Younsi_2012}:
\begin{equation}
    \frac{1}{\mathcal{C}} \frac{d}{d\lambda}\left(\frac{I_\nu}{\nu^3}\right) = \left(\frac{j_\nu}{\nu^2}\right) - (\nu \alpha_{\nu}) \left(\frac{I_\nu}{\nu^3}\right),
    \label{eq:radiative_transfer}
\end{equation}
where $\alpha_{\nu}$ is the frequency dependent absorption coefficient evaluated in the fluid frame. For thermal synchrotron, the absorption coefficient is defined as $\alpha_\nu = j_\nu/B_\nu$, where $B_\nu$ is the Planck function. We use the constant $\mathcal{C} \equiv  h \ell/m_e c^2$ in units of $\rm cm \ s$ to convert $d\lambda$ from code to physical units, where $\ell$ is the length unit and $h$ is Planck's constant in CGS. The specific intensity is proportional to the superphoton weight and is defined as
\begin{equation}
    I_\nu = \frac{h\nu dN}{dA \, dt \, d\nu \, d\Omega} = \frac{h \nu w dN_{\rm s}}{dA \, dt \, d\nu \, d\Omega} \propto w.
\end{equation}

At each step of the geodesic integration, absorption and scattering are taken into account as described in the following sections.

\subsection{Absorption}
\code{GPUmonty} handles absorption by decreasing the weight $w$ of the superphotons at each step of the geodesic integration as they travel through a medium. From the radiative transfer equation (Eq.~\ref{eq:radiative_transfer}) and ignoring emission ($j_\nu = 0$), the evolution of the superphoton weight due to absorption is given by
\begin{equation}
    \frac{dw}{d\tau_{\rm a}} = - w;
    \label{eq:weight_absorption}
\end{equation}
where the optical depth for absorption is $    d\tau_a = \nu \alpha_\nu \mathcal{C} d\lambda$. Integrating Equation~\ref{eq:weight_absorption}, the weight of each superphoton evolves as 
\begin{equation}
    w_{n + 1} = w_n e^{- \tau_a}.
\end{equation}

To compute the absorption optical depth over a single step, we evaluate the invariant absorption coefficient at the previous and current positions and approximate the integral of $d\tau$ over $d\lambda$ using the midpoint rule, following
\begin{equation}
    \tau_a = \frac{1}{2} \left[ (\nu \alpha_\nu)_n + (\nu \alpha_\nu)_{n + 1} \right] \mathcal{C} \Delta \lambda,
    \label{eq:abs_optdepth}
\end{equation}

\subsection{Scattering}
As superphotons propagate through the medium, we also account for potential scatterings. We model these using the probability distribution 
\begin{equation}
    p = 1 - e^{-b \tau_s}
\end{equation}
where $\tau_s$ represents the scattering optical depth and $b \geq 1$ is a numerical bias parameter used as a variance reduction technique to artificially improve the statistical sampling of scattering events, as discussed in \cite{kahn_1950, Dolence_2009}. The inclusion of $b$ is essential for enhancing the number of superphotons that experience scattering in optically thin scenarios, increasing the signal-to-noise ratio in scattering-dominated situations. When a scattering happens, a new superphoton is generated and we conserve the number of photons by setting the weight of the existing photon $w \rightarrow w(1 - 1/b)$ and the weight of the generated superphoton as $w/b$, such that $w (1 - 1/b) + w/b = 1$.

We account for the rate of interactions between photons and particles of mass $m$ by taking into account the cross-section invariance (equation 12.7 in \citealt{Landau_1975}; \citealt{Dolence_2009}) 
\begin{equation}
    \frac{1}{\sqrt{-g}} \frac{dN_{m\gamma}}{d^3x dt} = \int \frac{d^3p}{\sqrt{-g} p^t} \frac{dn_m}{d^3p} \frac{(- k_\mu p^\mu)}{k^t} \sigma c
\end{equation}
where $\sigma$ is the cross section and $dn_m = dN_m/d^3x$, where $dN_{m\gamma}$ is the number of interaction events (scatterings) between the mass particles and photons, while $dN_m$ is the number of particles contained within an infinitesimal volume $d^3x$. It is possible to define a ``hot cross section'' as
\begin{equation}
    \sigma_h = \frac{1}{n_m} \int d^3p \frac{dn_m}{d^3p} (1 - \mu_m \beta_m) \sigma,
\end{equation}
where $\beta$ is the particle speed, $\sigma$ is the invariant cross section and $\mu_m$ is the cosine of the angle between particle and photon momentum in the fluid frame (all quantities defined in the fluid frame). The extinction coefficient is 
\begin{equation}
    \alpha_\nu^{\rm sc} = n_m \sigma_h. 
\end{equation}
Hence, we calculate the scattering optical depth analogously to the absorption optical depth in equation \ref{eq:abs_optdepth}
\begin{equation}
    \tau_{\rm sc} = \frac{1}{2}\left[ (\nu \alpha_\nu^{\rm sc})_n + (\nu \alpha_{\nu}^{\rm sc})_{ n+1} \right] \mathcal{C} \Delta \lambda.
\end{equation}

In \code{GPUmonty}, $\sigma_h$ is calculated a priori and stored in a lookup table. We adopt the Klein-Nishina total cross section,

\begin{equation}
\sigma_{\rm KN} =
\sigma_T \frac{3}{4 \epsilon_e^2}
\left(
2 + \frac{\epsilon_e^2 (1 + \epsilon_e)}{(1 + 2\epsilon_e)^2}
+ \frac{\epsilon_e^2 - 2\epsilon_e - 2}{2\epsilon_e}
\log(1 + 2\epsilon_e)
\right)
\end{equation}
where $\sigma_T$ is the Thomson cross section. For $\epsilon \ll 1$, we adopt
\begin{equation}
\sigma_{\rm KN} \approx \sigma_T (1 - 2\epsilon).
\end{equation}
Numerically, to account for a scattering, we draw a random number $x_1 = - \log(\rm{rand})$, calculate the bias parameter and then check if
$b d\tau_{\rm{scat}} > x_1$ is satisfied. If this is the case, the scattering takes place.

A plasma-frame orthonormal tetrad is constructed via a Gram–Schmidt orthogonalization procedure, and the wave vector of the incoming superphoton is transformed to this frame. The four-momentum of the scattering electron is sampled from the local electron distribution function using a rejection-sampling method following the approach implemented in \code{igrmonty} \citep{Wong:2022rqr}. \code{igrmonty} is a modern implementation of \code{grmonty} maintained by the University of Illinois\footnote{Available for download at \url{https://github.com/AFD-Illinois/igrmonty}.}. For each candidate electron–photon pair, a second rejection step to sample the scattered superphoton based on the differential Compton scattering cross section, using the Thomson limit at low photon energies and the full Klein–Nishina cross section otherwise. This procedure is equivalent to the prescription of \citet{Canfield_1987} for isotropic thermal (Maxwell–Jüttner) electron distributions, but is more general and can be straightforwardly extended to non-thermal distribution functions. The scattered photon wave vector is finally constructed in the electron rest frame and boosted back to the coordinate frame.

\section{Implementation of the GPU-Accelerated Algorithms}
\label{sec:gpu_algorithms}

\code{GPUmonty} is developed in CUDA/C, leveraging OpenMP for CPU tasks while primarily operating on the GPU. We use device link-time optimization (dlto) to optimize function calls, in-lining, and memory usage throughout the program.

We use the well-established CUDA library cuRAND to implement the pseudo-random number generator XORWOW, which boasts a period of $2^{192} - 2^{32}$. By ``period'' we mean the length of the sequence of random numbers that the generator can produce before it begins to repeat. In the case of XORWOW, the period is slightly less than $2^{192}$. 

\code{GPUmonty} uses five main kernels (global functions) for photon creation, sampling, tracking, scattering, and recording. CPU-side tasks are limited to creating tables and reading data from external files, such as plasma properties from GRMHD snapshots, computing the metric determinant for each zone when calculating the weight for photon emission, and generating the output spectrum. We also minimize data transfer between GPU and CPU memories, so that PCIe bandwidth is not a concern. Most data transfer occurs before the creation of superphotons, primarily involving the transfer of table values and global variables. After the superphotons are recorded for the final spectrum, their data is transferred back from the GPU to the CPU. 

We generate, sample and track all superphotons at the same time, instead of generating them one by one as in \code{grmonty}. This minimizes the usage of atomic operations to global variables and also favors the Single Instruction Multiple Threads (SIMT) behavior of GPUs where threads are grouped into warps, and the same instruction is dispatched to all threads in a warp.

For photon generation, we employ stride-based parallelism, in which each GPU thread processes multiple plasma zones. Each thread is assigned a unique starting zone index equal to its global thread index. The thread then iterates over the domain by incrementing this index by a fixed stride equal to the total number of threads launched in the kernel. In this way, thread~0 processes zones $0, N_{\rm threads}, 2N_{\rm threads}, \ldots$, thread~1 processes zones $1, 1+N_{\rm threads}, 1+2N_{\rm threads}, \ldots$, and so on, ensuring that all plasma zones are covered exactly once.

For photon sampling and tracking, we use a different parallelism method. In this approach, each photon is assigned to a dedicated thread. Instead of using a fixed increment for each thread, the remaining photons are dynamically allocated as threads finish processing their assigned photons. This strategy is more efficient than stride-based parallelism because the time required for each photon to complete varies due to the rejection sampling method for the sampling methods described in Sec.~\ref{sec:sampling_procedure}, as well as due to different geodesics for different photons. As a result, threads that complete their tasks early are immediately reassigned to new photons, avoiding idle time while other threads finish processing.

In \code{grmonty}, photon scattering is handled through a recursive procedure, which is typically not ideal for GPUs due to their limited stack size. To overcome this limitation, we implemented an array of structures to store the properties of the scattered photons. These photons are then processed only after all original photons have been tracked. The scattering kernel manages this subsequent processing, allowing for multiple layers of scattering. Once all scattering events have been completed and recorded, the resulting spectrum data are transferred from device to host memory, and the output is written to a binary file.

Handling all the superphotons simultaneously has the downside of memory limitations, as it requires keeping track of every photon simultaneously in GPU memory. To overcome this, we first evaluate the available GPU memory and then divide the total number of superphotons into manageable batches that are executed serially. Each batch only saves the contribution of photons that achieved the recording criterion (not the photons themselves), i.e. the quantities entering the final observables, such as the superphoton weight, luminosity, energy, accumulated optical depths, scattering counts, and the detector bin. Therefore, the memory usage across batches is small. This algorithm allows us to free the memory of the processed superphotons before starting a new batch.

\section{Tests}
\label{sec:tests}

In this section, we present the tests performed to validate our code. The three tests performed are: (1) optically thin synchrotron sphere, (2) scattering in an uniform synchrotron sphere, and (3) comparison between \code{igrmonty} and \code{GPUmonty} for a GRMHD simulation. 

For the tests, we quantify the relative error between our method and the baseline as
\begin{equation}
    \delta L = \frac{L^{\rm ref}_\nu  - L^{\code{ours}}_\nu}{L^{\rm ref}_{\nu}},
    \label{eq:relative_error}
\end{equation}
where ``ours'' and ``ref'' indicate \code{GPUmonty} and the corresponding baseline, respectively. 

The convergence parameter is defined as
\begin{equation}
    \epsilon_{\rm err} = \frac{1}{\Delta \log(\nu)} \int \frac{\lvert L^{\rm ref}_\nu  - L^{\code{ours}}_\nu \rvert}{L^{\rm ref}_{\nu}} {\rm d}\log \nu.
    \label{eq:conv_parameter}
\end{equation}
where log has base 10.

\subsection{Optically thin synchrotron sphere}
\label{sec:optically_thin_sphere}

We begin the tests with a spherical synchrotron-emitting cloud in the optically thin regime, in flat spacetime. In this setup, we consider a homogeneous, uniform spherical cloud with a radius of $R_{\rm sphere} =1\ \rm{cm}$. We consider a grid with internal coordinates as $x^\mu = [t, \log(r), \theta, \phi]$. The cloud consists of relativistic electrons with a dimensionless temperature $\Theta_{\rm e} = 100$ and an electron number density $n_e = 10^{13}\ \rm{cm}^{-3}$. A vertical uniform magnetic field is considered, $B^z = 1$ G, which in our coordinates translates to
\begin{align}
    B^t & =  0,    \\
    B^r &= B_0 \cos\theta/r^2,    \\
    B^\theta &= - B_0 \sin\theta/r,    \\
    B^\phi &= 0,
\end{align}
with $B_0 = 1\ G$. With the photon optical path and the angle between the electron velocity and the magnetic field as $L = 1\ {\rm cm}$ and $\pi/2$ respectively at $\nu = 10^9 \ \rm{Hz}$, we estimate the optical depth as $\tau = \alpha_\nu L \approx 10^{-4}$, confirming that the sphere is optically thin. 

A 2D grid is initialized with a resolution of $8192 \times 128 \times 1$, with a maximum radius of $r_{\rm out} = 10000 \ \rm{cm}$. A high resolution in the radial direction is necessary to smooth the boundaries of the sphere. The polar angle is closed within the interval $[0, \pi]$. We set up $r_{\rm max} = 3000\ \rm{cm}$ as the radius at which to stop tracking and save the superphoton, and divide the energy in 2500 energy bins of size $\ln(h\nu/m_e c^2) = 0.01$ to accurately represent the rapidly varying high-end tail of the spectrum.

The results are compared with the angle-averaged emissivity $\hat{\epsilon}_\nu$ integrated over the volume of the sphere. Because the emissivity values are the same for every cell within the sphere, we can write
\begin{equation}
    L^{\rm ref}_{\nu} = \int \sqrt{-g} \hat{\epsilon}_\nu dV = \frac{4 \pi R_{\rm sphere}^3}{3} \hat{\epsilon}_\nu.
    \label{eq:Lnu_analytic_thin_test}
\end{equation}

\begin{figure}
    \centering
    \includegraphics[width=\linewidth]{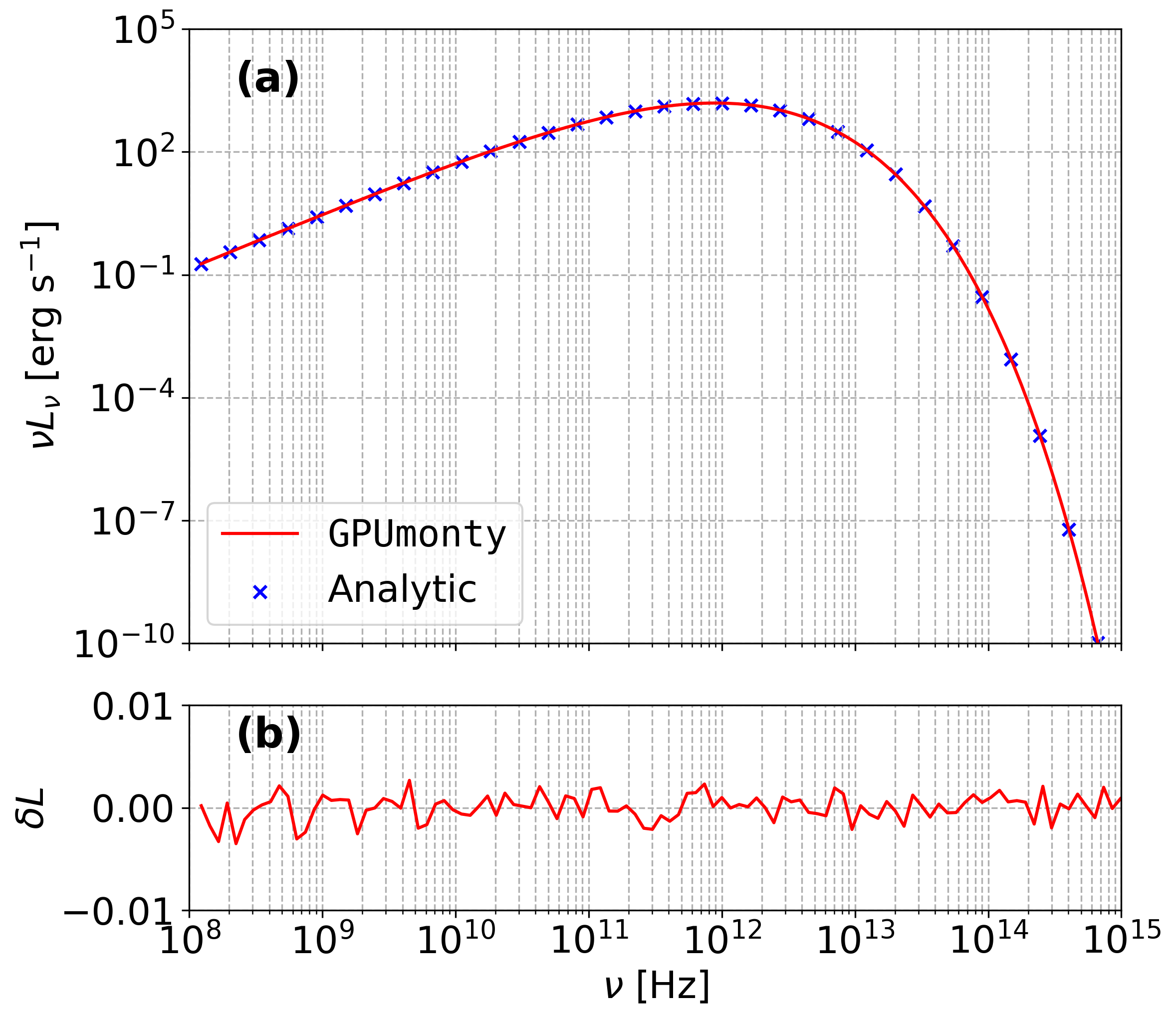}
    \caption{Panel~(a): Optically thin synchrotron sphere spectrum considering $N_{\rm s} = 10^8$. The red line represents the synthetic spectrum generated with \code{GPUmonty} and the purple markers are the analytical results from the angle-integrated emissivity calculated with equation \eqref{eq:Lnu_analytic_thin_test}. Panel~(b): Residuals computed with equation \eqref{eq:relative_error}. }
    \label{fig:thin_synch_test}
\end{figure}

Figure \ref{fig:thin_synch_test} compares the fiducial analytical and simulated spectra considering $N_{\rm s} = 10^8$. There is an excellent agreement between the numerical and analytical spectra, validated by the maximum difference between the numerical and analytical results remaining below $1\%$ in all energy bins. 

\begin{figure}
    \centering
    \includegraphics[width=\linewidth]{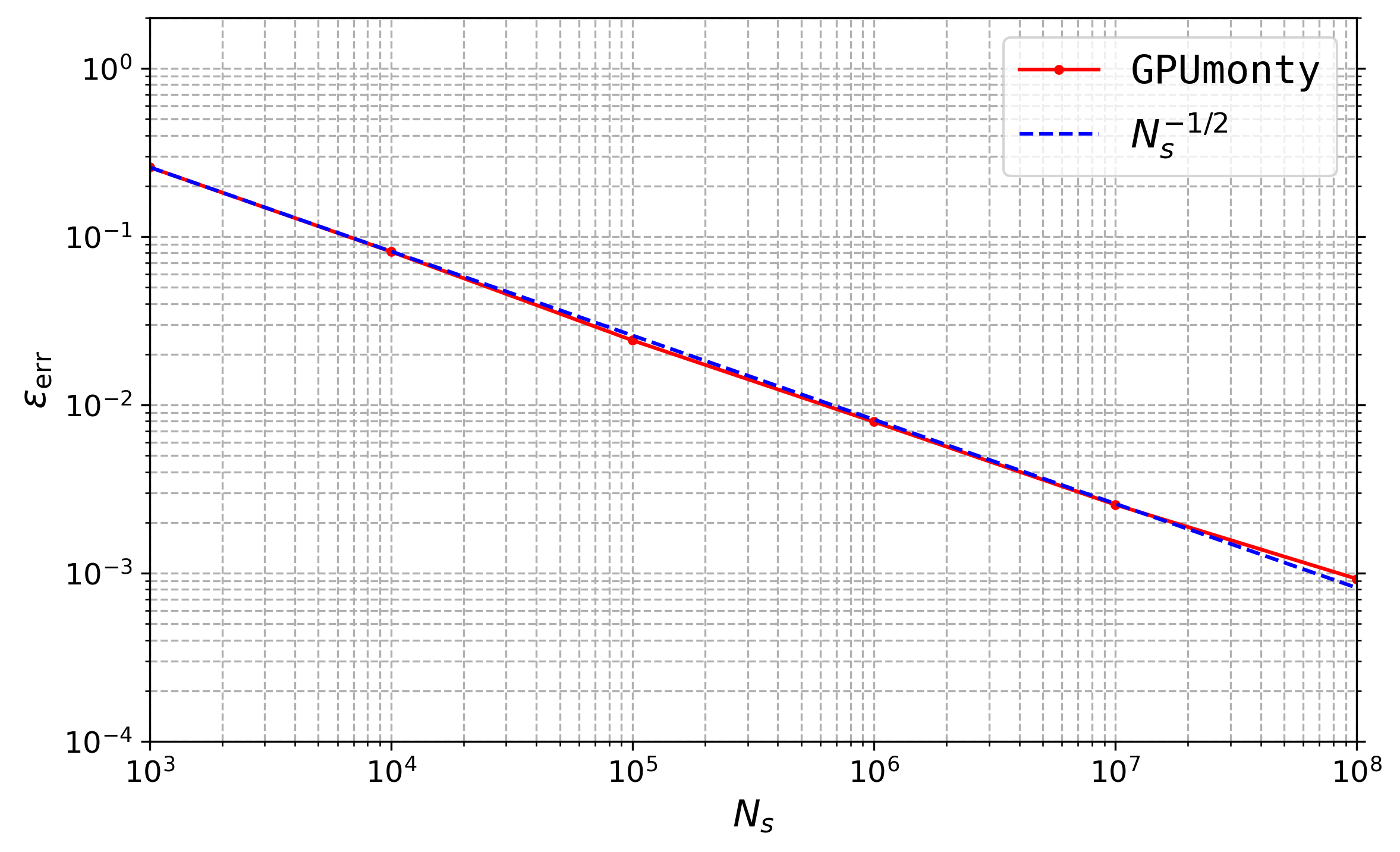}
    \caption{Normalized integrated error for different values of $N_{\rm s}$ in the optically thin sphere test, computed with equation~\eqref{eq:conv_parameter}. The blue dashed line represents the convergence $\epsilon_{\rm err} \propto N_{\rm s}^{-1/2}$ as expected in Monte Carlo simulations.}
    \label{fig:thin_synch_error}
\end{figure}

Figure \ref{fig:thin_synch_error} shows the normalized integrated error (eq. \ref{eq:relative_error}) as we vary the number of superphotons considering $N_{\rm s} = [10^4, 10^5, 10^6, 10^7, 10^8]$. The convergence scales proportional to $1/\sqrt{N_{\rm s}}$, as expected for the statistical error of a Monte Carlo estimator approaching the true underlying distribution. This result matches \code{igrmonty}'s.

We highlight that, at $10^8$ superphotons, small deviations from the ideal $1/\sqrt{N_{\rm s}}$ convergence arise due to discretization effects at the boundary of the sphere. Therefore, for $N_s = 10^8$ in Figure~\ref{fig:thin_synch_error}, we increase the radial resolution to $32000$ cells, reaching a grid spacing of $dx \sim 10^{-4}$. In addition, for this specific case, rather than imposing an abrupt cutoff of the gas variables at the sphere boundary, we adopt a partial-cell treatment. For cells whose edges extend beyond $R_{\rm sphere}$, we subdivide the cell into a $5\times5$ subgrid and scale the relevant variables by the fraction of the cell that lies within the sphere. The gas variables are then weighted by this fraction. For example, if $80\%$ of the subcells lie within $R_{\rm sphere}$, the electron density is set to $N_e = 0.8\,N_{e,\rm sphere}$.

\subsection{Scattering in an optically thick sphere}
\label{sec:scattering_test}

To validate the scattering algorithm, we consider the same uniform cloud of gas as described in Section~\ref{sec:optically_thin_sphere} but this time with an electron density $n_{\rm e} = 10^{18}\ \rm{cm^{-3}}$ and a dimensionless temperature of $\Theta_{\rm e} = 3$. Only in the scattering test, to enhance the scattering, the bias parameter is adopted as
\begin{equation}
b = n_{\rm e} \sigma_T R_{\rm sphere} (n_{\rm sc} + 1)^2,
\end{equation}
where $n_{\rm sc}$ indicates the scattering generation of the photon. In other words, photons with $n_{\rm sc} = 0$ are primary photons and the first scattered photons have $ n_\mathrm{sc} = 1$ etc. This choice of bias parameter ensures that we have sufficient photons to reduce statistical noise in the high-scattering regions. In an optically thin medium, as photons undergo successive scatterings, their number naturally decreases, so the bias parameter increases with scattering generation to enforce more scatterings and maintain adequate sampling of these higher layers. We adopt a heuristic square $(n_{\rm sc} + 1)^2$ scaling rather than a higher power to prevent the number of generated superphotons from becoming excessive, which could exceed available memory.  For this test, we allow for scatterings up to $n_{\rm sc} = 4$. 

\begin{figure}
    \centering
    \includegraphics[width=\linewidth]{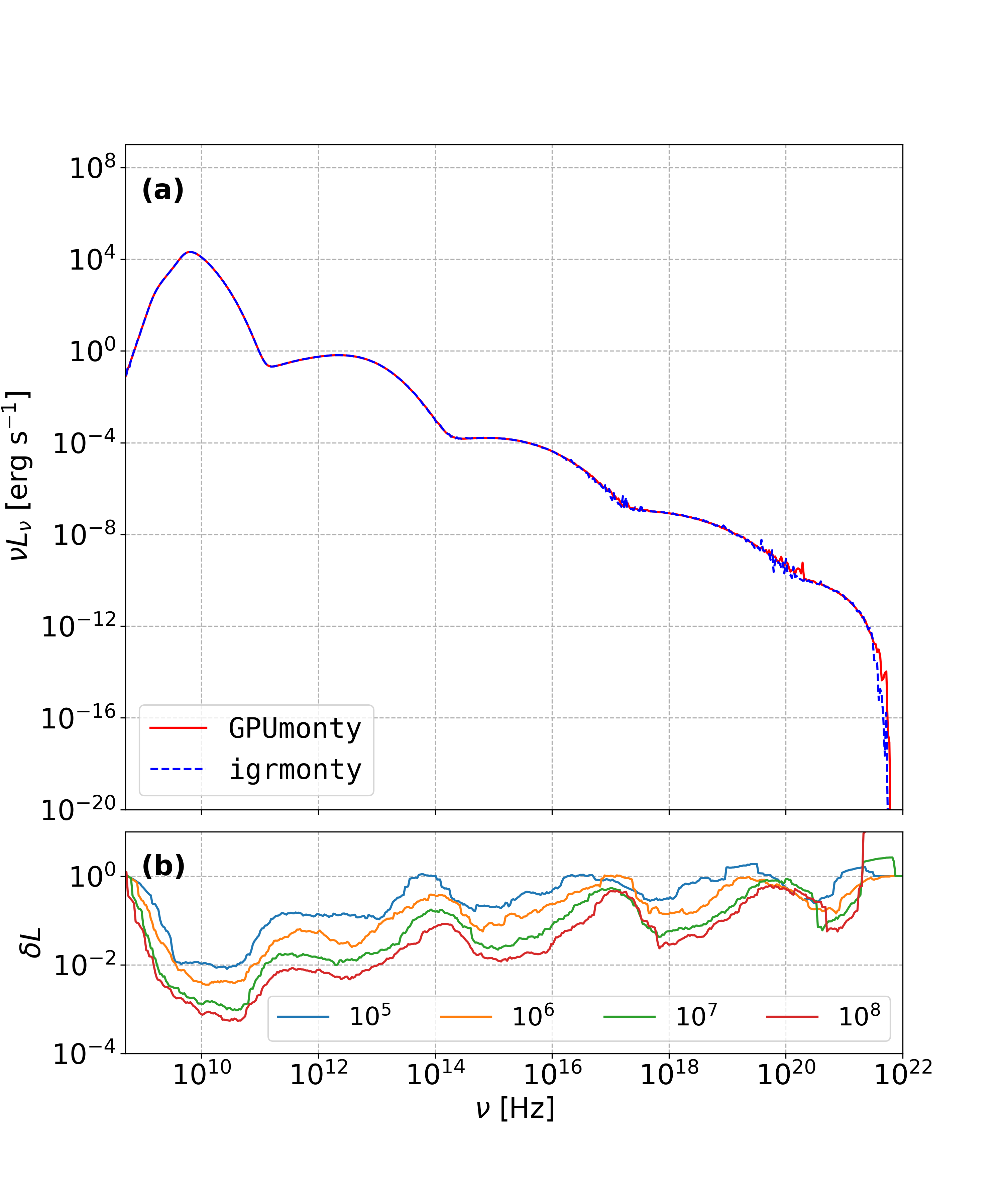}
    \caption{Panel (a): Scattering test for an uniformly spherical cloud of gas. The \code{GPUmonty} spectrum is shown with the red solid line and \code{igrmonty}'s with the blue dashed line, both with $N_{\rm s} = 10^8$.Panel (b): The frequency-dependent residuals, computed as in Figure \ref{fig:thin_synch_test}, for different values of $N_{\rm s}$. For clarity, the curves have been smoothed using a moving average over the $10$ nearest frequency bins.}
    \label{fig:scattering_comparison} 
\end{figure}

Figure~\ref{fig:scattering_comparison} shows the results from both \code{igrmonty} and \code{GPUmonty} for \( N_\mathrm{s} = 10^8 \) where each bump corresponds to a different scattering. There is an excellent agreement between the two methods. The tails of each bump exhibit increased noise which becomes more pronounced with higher \( n_\mathrm{sc} \). This elevated noise originates from the tail of the preceding bump, whose photons generally carry smaller weights, making it difficult to fully suppress noise in these regions. This effect gets intensified with successive scatterings.  As expected, as the number of superphotons increases the error decreases in the whole frequency domain except for the frequencies that match the end tail of each scattering round. The bottom panel of the figure shows the residuals where $N_\mathrm{S}$ is varied in \code{GPUmonty} and fix $N_\mathrm{S} = 10^8$ in \code{igrmonty}. 

\begin{figure}
    \centering
    \includegraphics[width=\linewidth]{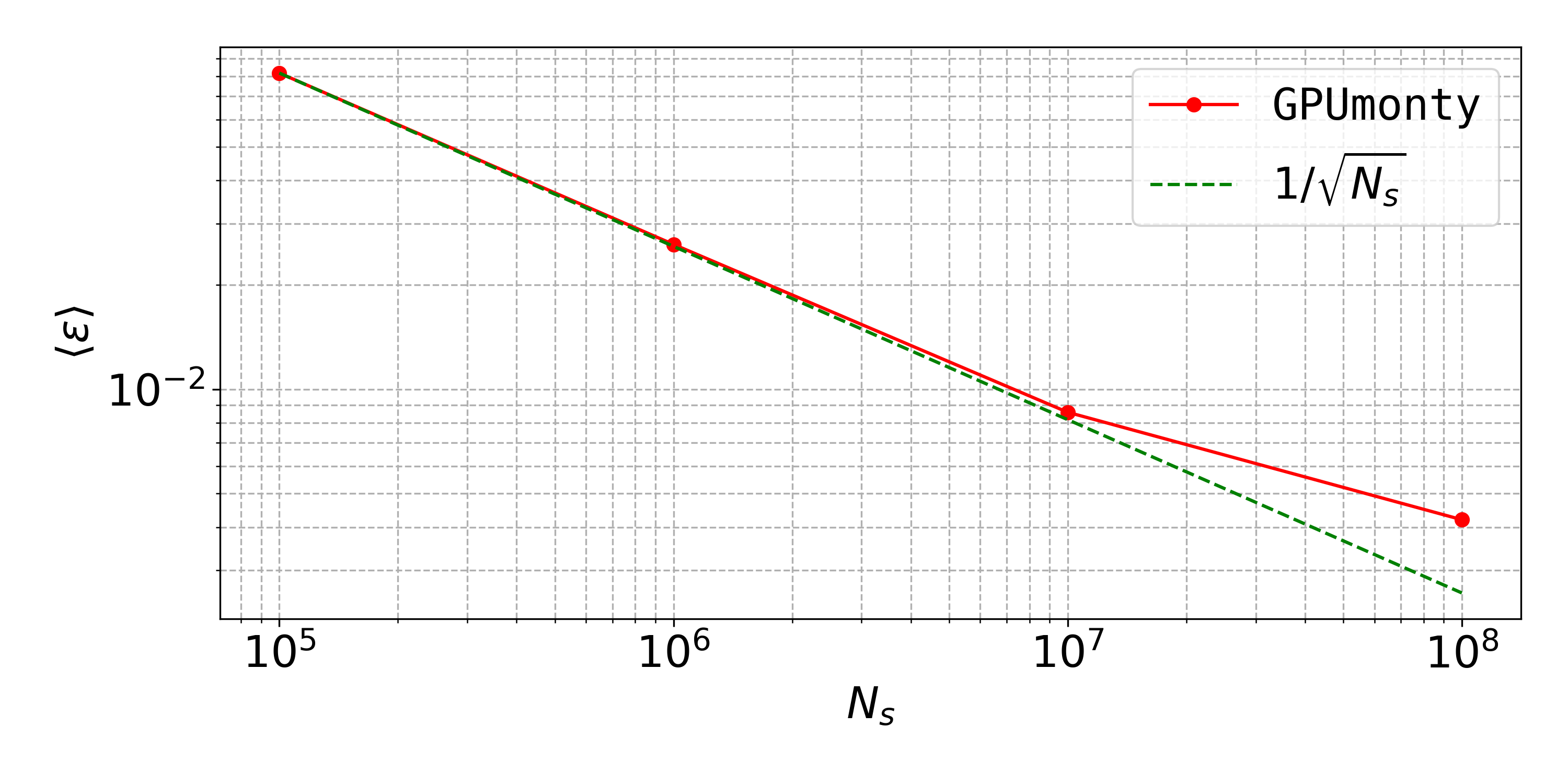}
    \caption{Normalized integrated error (equation~\eqref{eq:conv_parameter}) for different $N_{\rm s}$ sizes for the optically thick scattering test. The green dashed line represents $\epsilon_{\rm err} \propto N_{\rm s}^{-1/2}$.}
    \label{fig:scattering_comparison_error}
\end{figure}

Figure \ref{fig:scattering_comparison_error} shows the convergence parameter defined in Equation~\ref{eq:conv_parameter}. For this test, the integration is carried out up to $\nu = 10^{15}\ \rm{Hz}$, since the large errors observed in the higher-scattering bumps are also present in \code{igrmonty}, and would otherwise distort the convergence rate. Eliminating these errors would require running \code{igrmonty} with a significantly larger number of superphotons.

\subsection{GRMHD simulation} \label{sec:GRMHD_comparison}

In this section, we use the GRMHD code \code{iharm3d} \citep{Prather_2021} to simulate the accretion flow around a Kerr black hole. The snapshot used for the generation of the synthetic spectrum is the same one used for the EHT polarized radiative transfer code comparison \citep{Prather_2023}\footnote{The snapshot can be downloaded at \url{https://dataverse.harvard.edu/dataset.xhtml?persistentId=doi:10.7910/DVN/XZECPF}.}.

The simulation is conducted in 3D with a resolution of $288 \times 128 \times 128$ elements and begins with a \cite{Fishbone_1976} torus in Standard And Normal Evolution (SANE) magnetic topology \citep{Porth_2019} with a $a_\ast=0.9375$ spin. This simulation reflects the simulation dataset used in \citetalias{EHTM87Paper5} and \citet{EHTM87Paper7} (hereafter \citetalias{EHTM87Paper7}). The simulation details are further detailed in \citet{Wong:2022rqr}. The snapshot is taken at $4,500\ r_{\rm g}/c$, where $r_g$ is the gravitational radius given by $r_g = GM/c^2$, after the simulation starts, when the system reaches a quasi-steady accretion state in the inner disk regions. We consider a black hole of mass $M = 4.14 \times 10^6\ M_\odot$ as appropriate for Sgr A* and a conversion factor from code units to CGS of $\mathcal{M}= 1 \times 10^{16}\ \rm g$.

For our analysis, the minimum frequency is set at $10^8\ \rm Hz$ and the maximum frequency at $10^{16} \rm Hz$ for the synchrotron emission, with a minimum weight of $w = 10^{28}$. The energy bins are log-spaced, with a bin size defined by $\rm{ln}(h\nu/m_ec^2)=0.12$, accounting for a total of $800$ bins and starting the minimum energy bin at approximately $1.2 \times 10^8 \rm Hz$. The bias factor is set at $b=  96 \times 10^4 \times  \Theta_{\rm e}^2$. Finally, we consider $N_{\rm s} = 10^6$ and allow only photons that reach $r > 1000\, r_{\rm g}$ to contribute to the recorded spectrum.

\begin{figure}
    \centering
    \includegraphics[width=\linewidth]{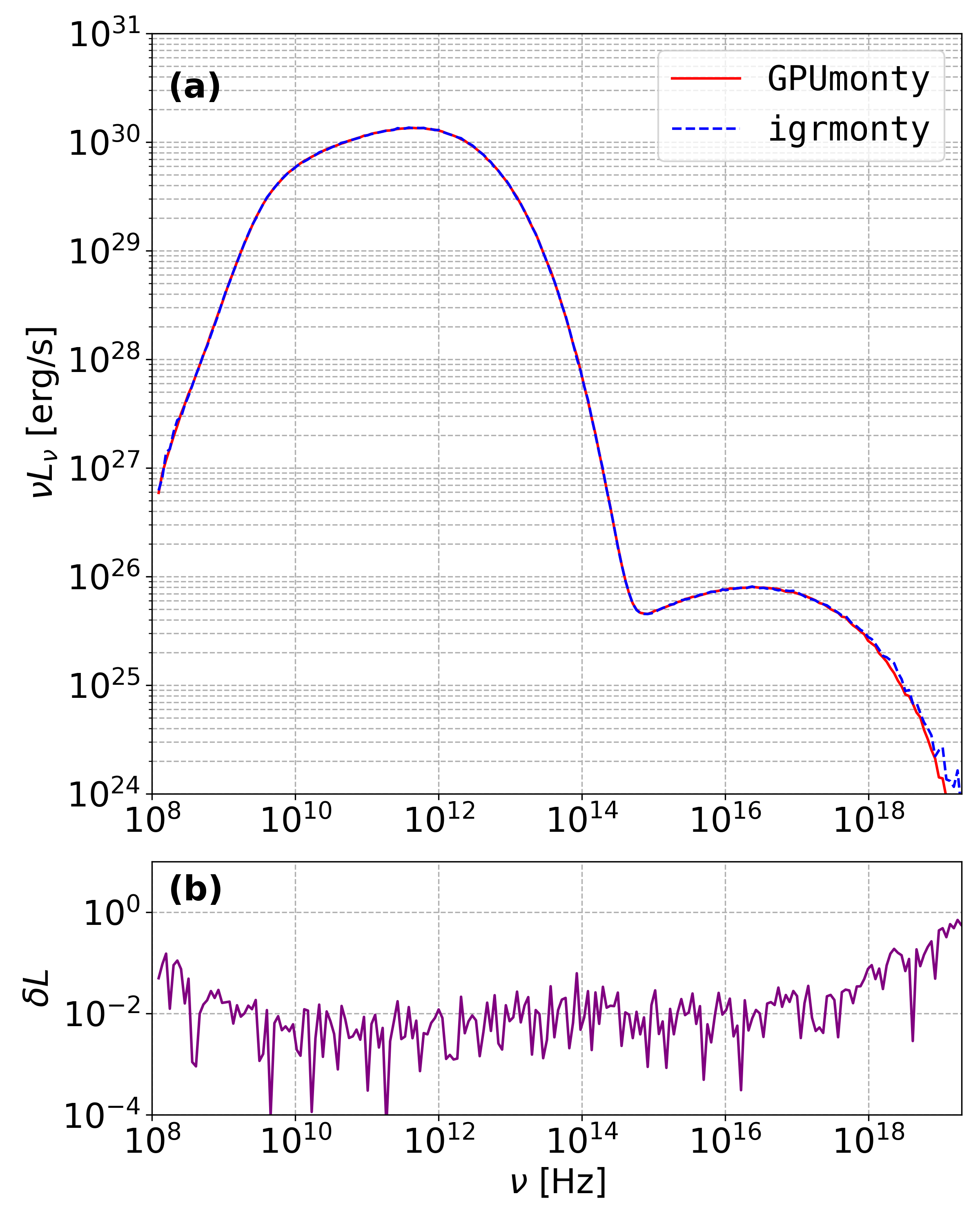}
    \caption{Panel~(a): Comparison of spectra computed with \code{GPUmonty} (red solid line) and \code{igrmonty} (blue dashed line) for a 3D SANE accretion flow GRMHD simulation around a Kerr black hole. Both spectra use the same parameters and $N_{\rm s} = 10^6$. Panel~(b): Residuals computed as in Figure \ref{fig:thin_synch_test} with \code{igrmonty} as baseline.}
    \label{fig:GRMHD_sane}
\end{figure}

Figure \ref{fig:GRMHD_sane} compares the spectra computed using the two radiative transfer codes for the selected GRMHD snapshot. The spectra agree closely over the full frequency range, with slight deviations only at the high-frequency tail, where the small number of scattered superphotons leads to Poisson noise. The error remains at the level of $\sim 10^{-2}$ across most frequencies and rises substantially only in the high-frequency tail for the aforementioned reason.

\section{GPU benchmark}
\label{sec:gpu_benchmark}

In this section, the performance of our GPU-accelerated algorithm is evaluated. We rely on NVIDIA's profiling tool \href{https://developer.nvidia.com/nsight-compute}{Nsight Compute}, which will provide all the metrics for this section.

The benchmarking is performed on the GRMHD simulation evaluated in Section \ref{sec:GRMHD_comparison}. The GPU runs were executed on an NVIDIA A100 GPU with 40~GB of HBM2 memory in the SXM form factor, featuring 108 streaming multiprocessors (SMs). CPU runs were performed on a single-socket AMD EPYC 7763 processor with 64 physical cores and 4 NUMA nodes. Each SM on the A100 can schedule up to 32 concurrent thread blocks, with a maximum of 1024 threads per block and a hardware limit of 2048 resident threads per SM. In our runs, we use 256 threads per block and a total number of blocks set to $108 \times 32 = 3456$, ensuring full occupancy of all SMs. Although \code{GPUmonty} is primarily GPU-driven, it relies on a few auxiliary CPU tasks as described in Section~\ref{sec:gpu_algorithms}. In the \code{GPUmonty} runs, we optimize CPU–GPU affinity and minimize NUMA-related overheads by allocating 16 OpenMP threads binding them to the NUMA node local to the GPU. In the \code{igrmonty} runs, we use a full dedicated AMD7763 node with the all the 64 cores.

\subsection{Runtime and speedup}

The performance of \code{GPUmonty} and \code{igrmonty} is analyzed as a function of the number of superphotons, spanning $N_{\rm s} = 10^3$–$10^9$. It is worth noting that the actual number of superphotons generated is usually larger than $N_{\rm s}$ by a factor of $\sim 7-10 \times$. All results presented here correspond to the same bias parameter value as chosen in Section~\ref{sec:GRMHD_comparison}, since modifying this parameter would change the number of scattered superphotons generated and thus alter the workload of the scattering routine. This could systematically impact the run times.

\begin{figure*}[htbp]
    \centering
    \includegraphics[width=0.8\linewidth]{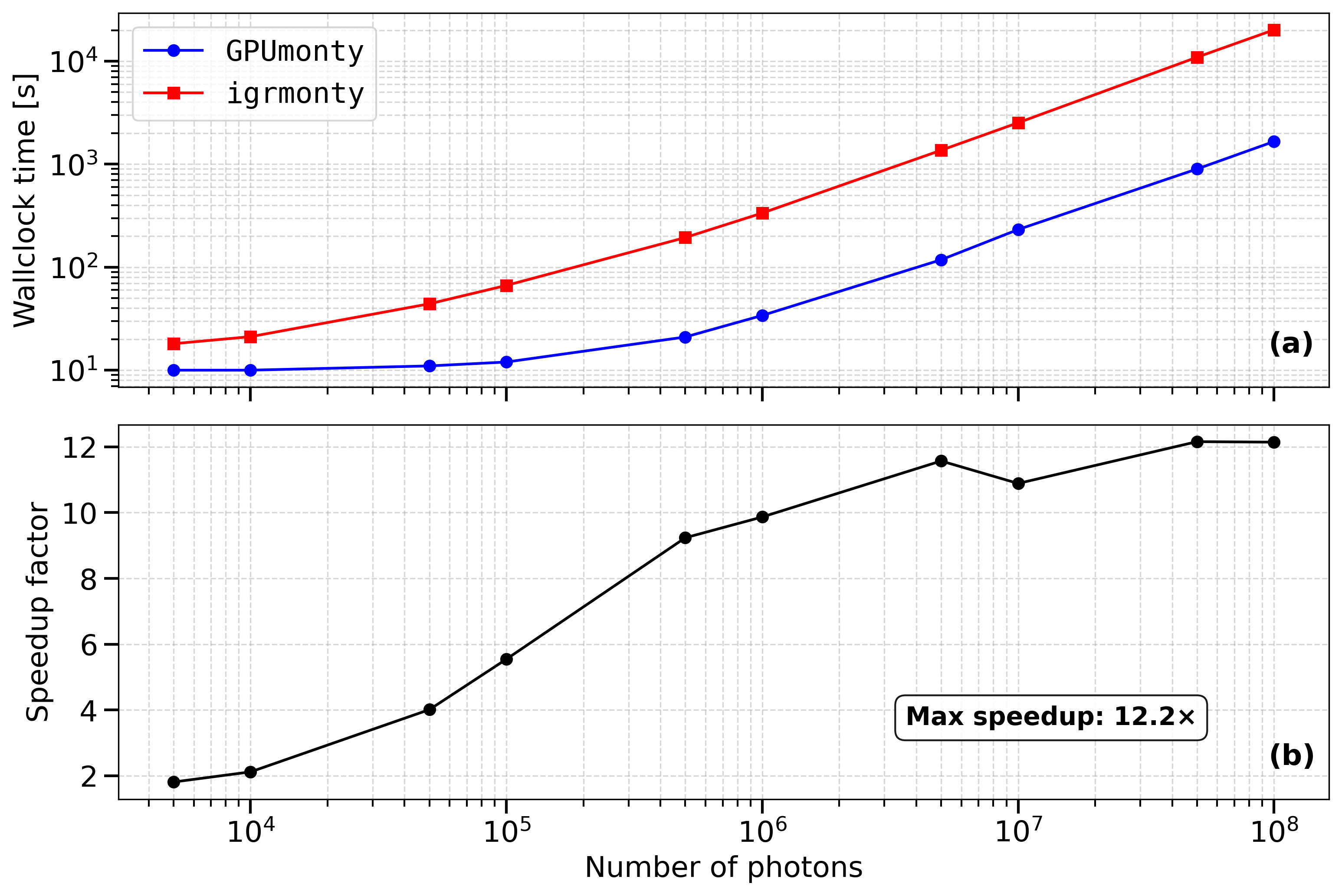}
    \caption{Panel~(a): Performance comparison between \code{GPUmonty} and \code{igrmonty} (CPU-based) as a function of superphoton number $N_{\rm s}$. Top panel: Wallclock time in seconds; \code{GPUmonty} in blue circles, \code{igrmonty} in red squares. Panel~(b): Resulting speedup factor achieved by \code{GPUmonty} relative to \code{igrmonty}. The speedup peaks at a factor of $\sim 12$ as the workload increases.}
    \label{fig:speedup_comparison}
\end{figure*}

The resulting execution times and speedup factors are depicted in Figure \ref{fig:speedup_comparison}.  We define the speedup factor as the ratio of \code{igrmonty} to \code{GPUmonty} execution time. Results are shown for a fixed bias parameter consistent with Section \ref{sec:GRMHD_comparison}. 

At low workloads corresponding to small photon numbers of $N_{\rm s} \lesssim 10^4$, the execution times of our GPU code and the pure CPU method are comparable. The reason is that for small $N_{\rm s}$, the overhead associated with GPU and CPU initialization and memory tasks (i.e. kernel launches, host-to-device data transfers; CPU-tasks such as data reading and table generation) is comparable to GPU computations related to photon generation and propagation. Once $N_{\rm s}$ increases, \code{GPUmonty} exhibits superior scaling compared to \code{igrmonty} due to our optimizations.

The speedup generally improves with the workload, reaching a maximum of approximately $12\times$. A slight drop in the speedup factor at $N_{\rm s} = 10^7$ is noticeable, which is explained by the workload exceeding the available GPU RAM, requiring the simulation to be processed in two serialized batches rather than a single pass. The batching serialization impacts the efficiency, though the speedup recovers as the workload increases further. Since we expect \code{GPUmonty} to typically run with $N_{\rm s} \gtrsim 10^6$ to ensure a high signal-to-noise ratio, \code{GPUmonty} effectively operates in the regime where the performance has already reached a plateau, providing a consistent speedup of approximately $12\, \times$.

We also quantify the codes' performances for the most efficient setup, corresponding to $N_{\rm s} = 5 \times 10^7$, by computing the number of processed superphotons per second. In the \code{GPUmonty} run, a total of $1{,}597{,}322{,}804$ superphotons are generated, including both plasma-emitted and scattered photons, over a runtime of $898$ seconds. This corresponds to a processing rate of approximately $1.78 \times 10^6$ superphotons/s. For the same value of $N_{\rm s}$, the \code{igrmonty} run produces $1{,}433{,}396{,}661$ superphotons with a total runtime of $10{,}906$ seconds, yielding a processing rate of $\sim 1.31 \times 10^5$ superphotons/s.

Performance on consumer-grade hardware was also evaluated to assess the speedup achievable on personal machines. The comparison was carried out between an NVIDIA GeForce RTX 3050 Laptop GPU (4 GB VRAM) and a AMD Ryzen 5 6600H, using $N_{\rm s} = 10^6$ in both cases. In this scenario, \code{GPUmonty} took $332$ seconds to run, while \code{igrmonty} took $2{,}692$ seconds, characterizing a solid speedup of $~\sim 8 \times$. As expected, the speedup is hardware-dependent and may vary with the specific system configuration.

\begin{table*}[htbp]
\centering
\small
\caption{GPU benchmarking metrics for different kernels in the simulation. Because some kernels are called more than one time, we report these parameters for the slowest execution of each kernel. The percentages enclosed by the parenthesis on the occupancy rows signals the theoretical occupancy for each kernel.}
\label{tab:gpu_benchmarks}

\setlength{\tabcolsep}{8pt}
\renewcommand{\arraystretch}{1.2} 

\begin{tabular}{lrrrrr}
\toprule
Metric & \multicolumn{1}{c}{Generation} & \multicolumn{1}{c}{Sampling} & \multicolumn{1}{c}{Tracking} & \multicolumn{1}{c}{Scattered} & \multicolumn{1}{c}{Recording} \\
\midrule
Kernel Duration              & 0.30\%& 0.20 \% & 42.32\% & 57.05\% & 0.13\%  \\
Achieved/Theoretical Occupancy & 95.2\% (12.5\%) & 99.7\% (12.5\%) & 100.0\% (12.5\%) & 100.0\% (12.5\%) & 95.2\% (62.5\%) \\
Compute Throughput           & 66.03\% & 57.99\% & 17.44\% & 17.98\% & 12.23\% \\
Memory Throughput            & 6.83\%  & 3.12\%  & 24.69\% & 26.13\% & 24.19\% \\

\bottomrule
\end{tabular}
\end{table*}

\subsection{Kernel execution times and bottlenecks}

For this analysis, the same setup is executed with $N_{\rm s} = 10^6$. Table~\ref{tab:gpu_benchmarks} reports metrics describing kernel execution time, compute throughput, memory and cache utilization, and SM occupancy, which together characterize how the kernel uses the GPU resources.

The kernel duration row shows that the total runtime is almost entirely dominated by the photon tracking stages. The superphoton tracking kernel alone accounts for $42\%$ of the execution time, while scattered tracking contributes an additional $57\%$. This indicates that performance optimization efforts should focus on the tracking stages, as improvements in other kernels would have a negligible impact on the overall runtime. The relative cost of the scattered tracking kernel depends on the chosen bias parameter. For the run shown here, this parameter leads to a larger number of scattered to generated superphotons.

An important factor limiting performance in our current implementation is the theoretical occupancy of the GPU, which measures the fraction of a streaming multiprocessor’s resources that can be actively used by warps given the kernel’s register and shared memory usage. In our case, the kernels are relatively long and require a substantial number of registers per thread to minimize register spilling, which inherently reduces the theoretical occupancy. In our tests, we found that reducing register pressure rather than maximizing occupancy actually led to shorter overall runtimes, making this trade-off beneficial for performance. While this means that not all hardware resources are fully used at all times, it allows each kernel to run efficiently without excessive memory access penalties. It is worth noting that this is an area where performance could be further optimized, and we plan to explore strategies to improve it in future work.

We find that prioritizing reduced register pressure over maximizing occupancy leads to improved performance. When forcing a lower register count (e.g. 32 registers per thread), the theoretical occupancy increases to $100\%$, with achieved occupancy exceeding $99.8\%$ for the tracking kernels. However, under this configuration the overall runtime increases compared to the baseline case with $12.5\%$ theoretical occupancy for most kernels, indicating that the increased register spilling and memory traffic outweigh the benefits of higher occupancy.

Achieving higher occupancy would require substantial refactoring of the code, potentially restructuring the algorithm into multiple smaller kernels. This may not be feasible given the inherently register-intensive nature of photon tracking. Despite these constraints, the current implementation still achieves a $12\times$ speedup relative to the reference CPU version, demonstrating that GPU acceleration remains highly effective. Future work will explore strategies to further reduce register usage, such as minimizing local variables, although the intrinsic complexity of the tracking algorithm may continue to impose high register demands.

\subsection{NVIDIA Nsight Compute metrics}
Compute throughput varies significantly across kernels. The superphoton generation and sampling kernels achieve high throughputs of $66\%$ and $58\%$ respectively, indicating that these kernels consist largely of regular arithmetic operations with limited branching and good instruction-level parallelism. In contrast, the tracking kernels exhibit significantly lower compute throughput, with $18\%$ for superphoton tracking and $22\%$ for scattered tracking. To determine if this low utilization was simply a byproduct of low occupancy ($12.5\%$), we performed a stress test by capping the register count at $32$ per thread.

This configuration achieved an occupancy of $99.9\%$, yet the compute throughput actually decreased to $\sim 10\%$. Simultaneously, memory throughput surged to $68.9\%$ due to massive register spilling to local memory. This result  demonstrates that even when the GPU is fully saturated with active warps, the compute pipes remain under-utilized. This confirms that the low compute throughput is intrinsic to the photon transport algorithm, which is dominated by complex control flow, instruction latency, and special-function units rather than long, throughput-oriented arithmetic sequences.

Memory throughput shows an inverse trend compared to compute throughput. The early-stage kernels make minimal use of global memory bandwidth, with values below $7\%$, reflecting their usage of register-resident data and limited memory traffic. In contrast, the tracking and recording kernels reach memory throughput values between $24\%$ and $31\%$. Although still well below saturation, these values indicate sustained interaction with global memory, driven by photon state updates, global spectrum variables recording and scattering bookkeeping. Importantly, all kernels operate well below the maximum available bandwidth, demonstrating that \code{GPUmonty} is not limited by memory bandwidth and that global memory access does not constitute the primary performance bottleneck. In the scenario of $32$ register per thread limit, the tracking kernels reached a memory throughput of approximately $68\%$. While this represents a significant increase in data movement, it remained below the maximum theoretical bandwidth of the A100.

\section{Conclusions}
\label{sec:conclusion}

This paper introduces the GPU-accelerated general relativistic Monte Carlo radiative transfer code \code{GPUmonty}. It is designed to compute the electromagnetic spectra emitted by hot gas in accretion flows around black holes due to the synchrotron and inverse Compton scattering processes. \code{GPUmonty} is a complete CUDA port of \code{grmonty}, making efficient use of NVIDIA GPUs via parallelism in five kernels: superphoton generation, sampling, tracking, scattering and recording. The original functions that relied on recursion, such as geodesic calculation and photon scattering, were restructured and optimized to accommodate the low stack size of GPUs.

\code{GPUmonty} is validated using three complementary tests designed to assess the accuracy of photon emission, absorption, and scattering: an optically thin synchrotron-emitting sphere and a self-synchrotron-Compton sphere in Minkowski spacetime, and a GRMHD simulation of a SANE radiatively inefficient accretion flow around a rapidly rotating black hole. Our benchmarks are, respectively, the corresponding analytical solution for the first test, and a \code{igrmonty} simulation for the second and third ones. In all tests, an excellent agreement with the benchmark is found, validating the physical accuracy and  expected Monte Carlo method statistical convergence of our GPU implementation. In our tests based on a GRMHD simulation,a $12 \times$ speedup is achieved using a single GPU when compared to \code{igrmonty}, a CPU-based fork of \code{grmonty}, on one 64-core CPU node.

In future work, \code{GPUmonty} will be extended with additional physical processes, including bremsstrahlung emission, non-thermal electron distributions, and photon polarization. Future updates will also include a bias-tuning process, where we will be able to control the amount of scattered photons to mitigate noise on the fly. The explicit separation of scattered-photon tracking in \code{GPUmonty} is expected to enable the implementation of new bias-tuning algorithms compared to those employed in \code{igrmonty}. On the performance side, optimization opportunities identified in Section~\ref{sec:gpu_benchmark} will be further investigated. Particular emphasis will be placed on mitigating register pressure through kernel reorganization. Portability to non-NVIDIA GPU architectures using HIP-based programming models will also be explored to broaden hardware support.

\section*{Acknowledgements}

The authors thank the anonymous referee for helpful comments and suggestions that improved the manuscript. PNM thanks Alejandro C\'ardenas-Avenda\~no, Douglas Ferreira, Trevor Gravely and Ben Prather for useful discussions, Angelina Lesniak for assistance with the algorithmic development, Reinaldo Lima for help with the testing algorithms, and Charles Gammie for valuable comments and suggestions on the manuscript. RN thanks Matheus T. Bernardino and Alfredo Goldman for their work on an early phase of this project. PNM also acknowledges the Center for Nonlinear Studies (CNLS) at Los Alamos National Laboratory (LANL) for hosting discussions that contributed to this work. RN gratefully acknowledges Rafa Munoz for the generous gift of a GPU, used in this research. PNM acknowledges financial support from the Fundação de Amparo à Pesquisa do Estado de São Paulo (FAPESP) under grant number 2023/15835-2. RN acknowledges a Bolsa de Produtividade from Conselho Nacional de Desenvolvimento Cient\'ifico e Tecnol\'ogico. This work used Delta CPU and GPU resources at the National Center for Supercomputing Applications (NCSA) through allocations PHY250391, PHY250091 and AST170024 from the Advanced Cyberinfrastructure Coordination Ecosystem: Services \& Support (ACCESS) program, which is supported by U.S. National Science Foundation grants \#2138259, \#2138286, \#2138307, \#2137603, and \#2138296. 

\section*{Data Availability}

\code{GPUmonty} is available on GitHub at \url{https://github.com/black-hole-group/gpumonty}, with documentation hosted at \url{https://black-hole-group.github.io/gpumonty/}. The code is also preserved on
Zenodo \citep{naethe_motta_2026_zenodo}. The project is released under the GNU GPL v2.0 license.

\bibliography{refs}{}

@article{Kawashima2023,
	adsurl = {https://ui.adsabs.harvard.edu/abs/2023ApJ...949..101K},
	archiveprefix = {arXiv},
	author = {{Kawashima}, Tomohisa and {Ohsuga}, Ken and {Takahashi}, Hiroyuki R.},
	doi = {10.3847/1538-4357/acc94a},
	eid = {101},
	eprint = {2108.05131},
	journal = {\apj},
	month = jun,
	number = {2},
	pages = {101},
	primaryclass = {astro-ph.HE},
	title = {{RAIKOU (来光): A General Relativistic, Multiwavelength Radiative Transfer Code}},
	volume = {949},
	year = 2023}

@article{Davelaar2023,
	adsurl = {https://ui.adsabs.harvard.edu/abs/2023MNRAS.526.5326D},
	archiveprefix = {arXiv},
	author = {{Davelaar}, Jordy and {Ryan}, Benjamin R. and {Wong}, George N. and {Bronzwaer}, Thomas and {Olivares}, Hector and {Mo{\'s}cibrodzka}, Monika and {Gammie}, Charles F. and {Falcke}, Heino},
	doi = {10.1093/mnras/stad3023},
	eprint = {2303.15522},
	journal = {\mnras},
	month = dec,
	number = {4},
	pages = {5326-5336},
	primaryclass = {astro-ph.HE},
	title = {{{\ensuremath{\kappa}}monty: a Monte Carlo Compton scattering code including non-thermal electrons}},
	volume = {526},
	year = 2023}

@article{Schnittman2013,
	adsurl = {http://adsabs.harvard.edu/abs/2013ApJ...777...11S},
	archiveprefix = {arXiv},
	arxivurl = {http://arXiv.org/abs/1302.3214},
	author = {{Schnittman}, J.~D. and {Krolik}, J.~H.},
	doi = {10.1088/0004-637X/777/1/11},
	eid = {11},
	eprint = {1302.3214},
	journal = {\apj},
	month = nov,
	pages = {11},
	primaryclass = {astro-ph.HE},
	title = {{A Monte Carlo Code for Relativistic Radiation Transport around Kerr Black Holes}},
	volume = 777,
	year = 2013}

@article{Chan2013,
	adsurl = {http://adsabs.harvard.edu/abs/2013ApJ...777...13C},
	archiveprefix = {arXiv},
	arxivurl = {http://arXiv.org/abs/1303.5057},
	author = {{Chan}, C.-k. and {Psaltis}, D. and {{\"O}zel}, F.},
	doi = {10.1088/0004-637X/777/1/13},
	eid = {13},
	eprint = {1303.5057},
	journal = {\apj},
	month = nov,
	pages = {13},
	primaryclass = {astro-ph.IM},
	title = {{GRay: A Massively Parallel GPU-based Code for Ray Tracing in Relativistic Spacetimes}},
	volume = 777,
	year = 2013}

@article{EHTC2019,
	adsurl = {https://ui.adsabs.harvard.edu/abs/2019ApJ...875L...1E},
	author = {{Event Horizon Telescope Collaboration} and {Akiyama}, Kazunori and {Alberdi}, Antxon and {Alef}, Walter and {Asada}, Keiichi and {Azulay}, Rebecca and {Baczko}, Anne-Kathrin and {Ball}, David and {Balokovi{\'c}}, Mislav and {Barrett}, John and {Bintley}, Dan and {Blackburn}, Lindy and {Boland}, Wilfred and {Bouman}, Katherine L. and {Bower}, Geoffrey C. and {Bremer}, Michael and {Brinkerink}, Christiaan D. and {Brissenden}, Roger and {Britzen}, Silke and {Broderick}, Avery E. and {Broguiere}, Dominique and {Bronzwaer}, Thomas and {Byun}, Do-Young and {Carlstrom}, John E. and {Chael}, Andrew and {Chan}, Chi-kwan and {Chatterjee}, Shami and {Chatterjee}, Koushik and {Chen}, Ming-Tang and {Chen}, Yongjun and {Cho}, Ilje and {Christian}, Pierre and {Conway}, John E. and {Cordes}, James M. and {Crew}, Geoffrey B. and {Cui}, Yuzhu and {Davelaar}, Jordy and {De Laurentis}, Mariafelicia and {Deane}, Roger and {Dempsey}, Jessica and {Desvignes}, Gregory and {Dexter}, Jason and {Doeleman}, Sheperd S. and {Eatough}, Ralph P. and {Falcke}, Heino and {Fish}, Vincent L. and {Fomalont}, Ed and {Fraga-Encinas}, Raquel and {Freeman}, William T. and {Friberg}, Per and {Fromm}, Christian M. and {G{\'o}mez}, Jos{\'e} L. and {Galison}, Peter and {Gammie}, Charles F. and {Garc{\'\i}a}, Roberto and {Gentaz}, Olivier and {Georgiev}, Boris and {Goddi}, Ciriaco and {Gold}, Roman and {Gu}, Minfeng and {Gurwell}, Mark and {Hada}, Kazuhiro and {Hecht}, Michael H. and {Hesper}, Ronald and {Ho}, Luis C. and {Ho}, Paul and {Honma}, Mareki and {Huang}, Chih-Wei L. and {Huang}, Lei and {Hughes}, David H. and {Ikeda}, Shiro and {Inoue}, Makoto and {Issaoun}, Sara and {James}, David J. and {Jannuzi}, Buell T. and {Janssen}, Michael and {Jeter}, Britton and {Jiang}, Wu and {Johnson}, Michael D. and {Jorstad}, Svetlana and {Jung}, Taehyun and {Karami}, Mansour and {Karuppusamy}, Ramesh and {Kawashima}, Tomohisa and {Keating}, Garrett K. and {Kettenis}, Mark and {Kim}, Jae-Young and {Kim}, Junhan and {Kim}, Jongsoo and {Kino}, Motoki and {Koay}, Jun Yi and {Koch}, Patrick M. and {Koyama}, Shoko and {Kramer}, Michael and {Kramer}, Carsten and {Krichbaum}, Thomas P. and {Kuo}, Cheng-Yu and {Lauer}, Tod R. and {Lee}, Sang-Sung and {Li}, Yan-Rong and {Li}, Zhiyuan and {Lindqvist}, Michael and {Liu}, Kuo and {Liuzzo}, Elisabetta and {Lo}, Wen-Ping and {Lobanov}, Andrei P. and {Loinard}, Laurent and {Lonsdale}, Colin and {Lu}, Ru-Sen and {MacDonald}, Nicholas R. and {Mao}, Jirong and {Markoff}, Sera and {Marrone}, Daniel P. and {Marscher}, Alan P. and {Mart{\'\i}-Vidal}, Iv{\'a}n and {Matsushita}, Satoki and {Matthews}, Lynn D. and {Medeiros}, Lia and {Menten}, Karl M. and {Mizuno}, Yosuke and {Mizuno}, Izumi and {Moran}, James M. and {Moriyama}, Kotaro and {Moscibrodzka}, Monika and {M{\"u}ller}, Cornelia and {Nagai}, Hiroshi and {Nagar}, Neil M. and {Nakamura}, Masanori and {Narayan}, Ramesh and {Narayanan}, Gopal and {Natarajan}, Iniyan and {Neri}, Roberto and {Ni}, Chunchong and {Noutsos}, Aristeidis and {Okino}, Hiroki and {Olivares}, H{\'e}ctor and {Ortiz-Le{\'o}n}, Gisela N. and {Oyama}, Tomoaki and {{\"O}zel}, Feryal and {Palumbo}, Daniel C.~M. and {Patel}, Nimesh and {Pen}, Ue-Li and {Pesce}, Dominic W. and {Pi{\'e}tu}, Vincent and {Plambeck}, Richard and {PopStefanija}, Aleksandar and {Porth}, Oliver and {Prather}, Ben and {Preciado-L{\'o}pez}, Jorge A. and {Psaltis}, Dimitrios and {Pu}, Hung-Yi and {Ramakrishnan}, Venkatessh and {Rao}, Ramprasad and {Rawlings}, Mark G. and {Raymond}, Alexander W. and {Rezzolla}, Luciano and {Ripperda}, Bart and {Roelofs}, Freek and {Rogers}, Alan and {Ros}, Eduardo and {Rose}, Mel and {Roshanineshat}, Arash and {Rottmann}, Helge and {Roy}, Alan L. and {Ruszczyk}, Chet and {Ryan}, Benjamin R. and {Rygl}, Kazi L.~J. and {S{\'a}nchez}, Salvador and {S{\'a}nchez-Arguelles}, David and {Sasada}, Mahito and {Savolainen}, Tuomas and {Schloerb}, F. Peter and {Schuster}, Karl-Friedrich and {Shao}, Lijing and {Shen}, Zhiqiang and {Small}, Des and {Sohn}, Bong Won and {SooHoo}, Jason and {Tazaki}, Fumie and {Tiede}, Paul and {Tilanus}, Remo P.~J. and {Titus}, Michael and {Toma}, Kenji and {Torne}, Pablo and {Trent}, Tyler and {Trippe}, Sascha and {Tsuda}, Shuichiro and {van Bemmel}, Ilse and {van Langevelde}, Huib Jan and {van Rossum}, Daniel R. and {Wagner}, Jan and {Wardle}, John and {Weintroub}, Jonathan and {Wex}, Norbert and {Wharton}, Robert and {Wielgus}, Maciek and {Wong}, George N. and {Wu}, Qingwen and {Young}, Ken and {Young}, Andr{\'e} and {Younsi}, Ziri and {Yuan}, Feng and {Yuan}, Ye-Fei and {Zensus}, J. Anton and {Zhao}, Guangyao and {Zhao}, Shan-Shan and {Zhu}, Ziyan and {Algaba}, Juan-Carlos and {Allardi}, Alexander and {Amestica}, Rodrigo and {Anczarski}, Jadyn and {Bach}, Uwe and {Baganoff}, Frederick K. and {Beaudoin}, Christopher and {Benson}, Bradford A. and {Berthold}, Ryan and {Blanchard}, Jay M. and {Blundell}, Ray and {Bustamente}, Sandra and {Cappallo}, Roger and {Castillo-Dom{\'\i}nguez}, Edgar and {Chang}, Chih-Cheng and {Chang}, Shu-Hao and {Chang}, Song-Chu and {Chen}, Chung-Chen and {Chilson}, Ryan and {Chuter}, Tim C. and {C{\'o}rdova Rosado}, Rodrigo and {Coulson}, Iain M. and {Crawford}, Thomas M. and {Crowley}, Joseph and {David}, John and {Derome}, Mark and {Dexter}, Matthew and {Dornbusch}, Sven and {Dudevoir}, Kevin A. and {Dzib}, Sergio A. and {Eckart}, Andreas and {Eckert}, Chris and {Erickson}, Neal R. and {Everett}, Wendeline B. and {Faber}, Aaron and {Farah}, Joseph R. and {Fath}, Vernon and {Folkers}, Thomas W. and {Forbes}, David C. and {Freund}, Robert and {G{\'o}mez-Ruiz}, Arturo I. and {Gale}, David M. and {Gao}, Feng and {Geertsema}, Gertie and {Graham}, David A. and {Greer}, Christopher H. and {Grosslein}, Ronald and {Gueth}, Fr{\'e}d{\'e}ric and {Haggard}, Daryl and {Halverson}, Nils W. and {Han}, Chih-Chiang and {Han}, Kuo-Chang and {Hao}, Jinchi and {Hasegawa}, Yutaka and {Henning}, Jason W. and {Hern{\'a}ndez-G{\'o}mez}, Antonio and {Herrero-Illana}, Rub{\'e}n and {Heyminck}, Stefan and {Hirota}, Akihiko and {Hoge}, James and {Huang}, Yau-De and {Impellizzeri}, C.~M. Violette and {Jiang}, Homin and {Kamble}, Atish and {Keisler}, Ryan and {Kimura}, Kimihiro and {Kono}, Yusuke and {Kubo}, Derek and {Kuroda}, John and {Lacasse}, Richard and {Laing}, Robert A. and {Leitch}, Erik M. and {Li}, Chao-Te and {Lin}, Lupin C. -C. and {Liu}, Ching-Tang and {Liu}, Kuan-Yu and {Lu}, Li-Ming and {Marson}, Ralph G. and {Martin-Cocher}, Pierre L. and {Massingill}, Kyle D. and {Matulonis}, Callie and {McColl}, Martin P. and {McWhirter}, Stephen R. and {Messias}, Hugo and {Meyer-Zhao}, Zheng and {Michalik}, Daniel and {Monta{\~n}a}, Alfredo and {Montgomerie}, William and {Mora-Klein}, Matias and {Muders}, Dirk and {Nadolski}, Andrew and {Navarro}, Santiago and {Neilsen}, Joseph and {Nguyen}, Chi H. and {Nishioka}, Hiroaki and {Norton}, Timothy and {Nowak}, Michael A. and {Nystrom}, George and {Ogawa}, Hideo and {Oshiro}, Peter and {Oyama}, Tomoaki and {Parsons}, Harriet and {Paine}, Scott N. and {Pe{\~n}alver}, Juan and {Phillips}, Neil M. and {Poirier}, Michael and {Pradel}, Nicolas and {Primiani}, Rurik A. and {Raffin}, Philippe A. and {Rahlin}, Alexandra S. and {Reiland}, George and {Risacher}, Christopher and {Ruiz}, Ignacio and {S{\'a}ez-Mada{\'\i}n}, Alejandro F. and {Sassella}, Remi and {Schellart}, Pim and {Shaw}, Paul and {Silva}, Kevin M. and {Shiokawa}, Hotaka and {Smith}, David R. and {Snow}, William and {Souccar}, Kamal and {Sousa}, Don and {Sridharan}, T.~K. and {Srinivasan}, Ranjani and {Stahm}, William and {Stark}, Anthony A. and {Story}, Kyle and {Timmer}, Sjoerd T. and {Vertatschitsch}, Laura and {Walther}, Craig and {Wei}, Ta-Shun and {Whitehorn}, Nathan and {Whitney}, Alan R. and {Woody}, David P. and {Wouterloot}, Jan G.~A. and {Wright}, Melvin and {Yamaguchi}, Paul and {Yu}, Chen-Yu and {Zeballos}, Milagros and {Zhang}, Shuo and {Ziurys}, Lucy},
	doi = {10.3847/2041-8213/ab0ec7},
	eid = {L1},
	journal = {\apj},
	month = {Apr},
	number = {1},
	pages = {L1},
	title = {{First M87 Event Horizon Telescope Results. I. The Shadow of the Supermassive Black Hole}},
	volume = {875},
	year = {2019}}

@software{naethe_motta_2026_zenodo,
  author       = {Naethe Motta, Pedro and
                  Nemmen, Rodrigo and
                  Joshi, Abhishek},
  title        = {GPUmonty},
  month        = feb,
  year         = 2026,
  publisher    = {Zenodo},
  doi          = {10.5281/zenodo.18884082},
  url          = {https://doi.org/10.5281/zenodo.18884082},
}

@article{EHTC2022,
	adsurl = {https://ui.adsabs.harvard.edu/abs/2022ApJ...930L..12E},
	author = {{Event Horizon Telescope Collaboration} and {Akiyama}, Kazunori and {Alberdi}, Antxon and {Alef}, Walter and {Algaba}, Juan Carlos and {Anantua}, Richard and {Asada}, Keiichi and {Azulay}, Rebecca and {Bach}, Uwe and {Baczko}, Anne-Kathrin and {Ball}, David and {Balokovi{\'c}}, Mislav and {Barrett}, John and {Baub{\"o}ck}, Michi and {Benson}, Bradford A. and {Bintley}, Dan and {Blackburn}, Lindy and {Blundell}, Raymond and {Bouman}, Katherine L. and {Bower}, Geoffrey C. and {Boyce}, Hope and {Bremer}, Michael and {Brinkerink}, Christiaan D. and {Brissenden}, Roger and {Britzen}, Silke and {Broderick}, Avery E. and {Broguiere}, Dominique and {Bronzwaer}, Thomas and {Bustamante}, Sandra and {Byun}, Do-Young and {Carlstrom}, John E. and {Ceccobello}, Chiara and {Chael}, Andrew and {Chan}, Chi-kwan and {Chatterjee}, Koushik and {Chatterjee}, Shami and {Chen}, Ming-Tang and {Chen}, Yongjun and {Cheng}, Xiaopeng and {Cho}, Ilje and {Christian}, Pierre and {Conroy}, Nicholas S. and {Conway}, John E. and {Cordes}, James M. and {Crawford}, Thomas M. and {Crew}, Geoffrey B. and {Cruz-Osorio}, Alejandro and {Cui}, Yuzhu and {Davelaar}, Jordy and {Laurentis}, Mariafelicia De and {Deane}, Roger and {Dempsey}, Jessica and {Desvignes}, Gregory and {Dexter}, Jason and {Dhruv}, Vedant and {Doeleman}, Sheperd S. and {Dougal}, Sean and {Dzib}, Sergio A. and {Eatough}, Ralph P. and {Emami}, Razieh and {Falcke}, Heino and {Farah}, Joseph and {Fish}, Vincent L. and {Fomalont}, Ed and {Ford}, H. Alyson and {Fraga-Encinas}, Raquel and {Freeman}, William T. and {Friberg}, Per and {Fromm}, Christian M. and {Fuentes}, Antonio and {Galison}, Peter and {Gammie}, Charles F. and {Garc{\'\i}a}, Roberto and {Gentaz}, Olivier and {Georgiev}, Boris and {Goddi}, Ciriaco and {Gold}, Roman and {G{\'o}mez-Ruiz}, Arturo I. and {G{\'o}mez}, Jos{\'e} L. and {Gu}, Minfeng and {Gurwell}, Mark and {Hada}, Kazuhiro and {Haggard}, Daryl and {Haworth}, Kari and {Hecht}, Michael H. and {Hesper}, Ronald and {Heumann}, Dirk and {Ho}, Luis C. and {Ho}, Paul and {Honma}, Mareki and {Huang}, Chih-Wei L. and {Huang}, Lei and {Hughes}, David H. and {Ikeda}, Shiro and {Impellizzeri}, C.~M. Violette and {Inoue}, Makoto and {Issaoun}, Sara and {James}, David J. and {Jannuzi}, Buell T. and {Janssen}, Michael and {Jeter}, Britton and {Jiang}, Wu and {Jim{\'e}nez-Rosales}, Alejandra and {Johnson}, Michael D. and {Jorstad}, Svetlana and {Joshi}, Abhishek V. and {Jung}, Taehyun and {Karami}, Mansour and {Karuppusamy}, Ramesh and {Kawashima}, Tomohisa and {Keating}, Garrett K. and {Kettenis}, Mark and {Kim}, Dong-Jin and {Kim}, Jae-Young and {Kim}, Jongsoo and {Kim}, Junhan and {Kino}, Motoki and {Koay}, Jun Yi and {Kocherlakota}, Prashant and {Kofuji}, Yutaro and {Koch}, Patrick M. and {Koyama}, Shoko and {Kramer}, Carsten and {Kramer}, Michael and {Krichbaum}, Thomas P. and {Kuo}, Cheng-Yu and {Bella}, Noemi La and {Lauer}, Tod R. and {Lee}, Daeyoung and {Lee}, Sang-Sung and {Leung}, Po Kin and {Levis}, Aviad and {Li}, Zhiyuan and {Lico}, Rocco and {Lindahl}, Greg and {Lindqvist}, Michael and {Lisakov}, Mikhail and {Liu}, Jun and {Liu}, Kuo and {Liuzzo}, Elisabetta and {Lo}, Wen-Ping and {Lobanov}, Andrei P. and {Loinard}, Laurent and {Lonsdale}, Colin J. and {Lu}, Ru-Sen and {Mao}, Jirong and {Marchili}, Nicola and {Markoff}, Sera and {Marrone}, Daniel P. and {Marscher}, Alan P. and {Mart{\'\i}-Vidal}, Iv{\'a}n and {Matsushita}, Satoki and {Matthews}, Lynn D. and {Medeiros}, Lia and {Menten}, Karl M. and {Michalik}, Daniel and {Mizuno}, Izumi and {Mizuno}, Yosuke and {Moran}, James M. and {Moriyama}, Kotaro and {Moscibrodzka}, Monika and {M{\"u}ller}, Cornelia and {Mus}, Alejandro and {Musoke}, Gibwa and {Myserlis}, Ioannis and {Nadolski}, Andrew and {Nagai}, Hiroshi and {Nagar}, Neil M. and {Nakamura}, Masanori and {Narayan}, Ramesh and {Narayanan}, Gopal and {Natarajan}, Iniyan and {Nathanail}, Antonios and {Fuentes}, Santiago Navarro and {Neilsen}, Joey and {Neri}, Roberto and {Ni}, Chunchong and {Noutsos}, Aristeidis and {Nowak}, Michael A. and {Oh}, Junghwan and {Okino}, Hiroki and {Olivares}, H{\'e}ctor and {Ortiz-Le{\'o}n}, Gisela N. and {Oyama}, Tomoaki and {{\"O}zel}, Feryal and {Palumbo}, Daniel C.~M. and {Paraschos}, Georgios Filippos and {Park}, Jongho and {Parsons}, Harriet and {Patel}, Nimesh and {Pen}, Ue-Li and {Pesce}, Dominic W. and {Pi{\'e}tu}, Vincent and {Plambeck}, Richard and {PopStefanija}, Aleksandar and {Porth}, Oliver and {P{\"o}tzl}, Felix M. and {Prather}, Ben and {Preciado-L{\'o}pez}, Jorge A. and {Psaltis}, Dimitrios and {Pu}, Hung-Yi and {Ramakrishnan}, Venkatessh and {Rao}, Ramprasad and {Rawlings}, Mark G. and {Raymond}, Alexander W. and {Rezzolla}, Luciano and {Ricarte}, Angelo and {Ripperda}, Bart and {Roelofs}, Freek and {Rogers}, Alan and {Ros}, Eduardo and {Romero-Ca{\~n}izales}, Cristina and {Roshanineshat}, Arash and {Rottmann}, Helge and {Roy}, Alan L. and {Ruiz}, Ignacio and {Ruszczyk}, Chet and {Rygl}, Kazi L.~J. and {S{\'a}nchez}, Salvador and {S{\'a}nchez-Arg{\"u}elles}, David and {S{\'a}nchez-Portal}, Miguel and {Sasada}, Mahito and {Satapathy}, Kaushik and {Savolainen}, Tuomas and {Schloerb}, F. Peter and {Schonfeld}, Jonathan and {Schuster}, Karl-Friedrich and {Shao}, Lijing and {Shen}, Zhiqiang and {Small}, Des and {Sohn}, Bong Won and {SooHoo}, Jason and {Souccar}, Kamal and {Sun}, He and {Tazaki}, Fumie and {Tetarenko}, Alexandra J. and {Tiede}, Paul and {Tilanus}, Remo P.~J. and {Titus}, Michael and {Torne}, Pablo and {Traianou}, Efthalia and {Trent}, Tyler and {Trippe}, Sascha and {Turk}, Matthew and {van Bemmel}, Ilse and {van Langevelde}, Huib Jan and {van Rossum}, Daniel R. and {Vos}, Jesse and {Wagner}, Jan and {Ward-Thompson}, Derek and {Wardle}, John and {Weintroub}, Jonathan and {Wex}, Norbert and {Wharton}, Robert and {Wielgus}, Maciek and {Wiik}, Kaj and {Witzel}, Gunther and {Wondrak}, Michael F. and {Wong}, George N. and {Wu}, Qingwen and {Yamaguchi}, Paul and {Yoon}, Doosoo and {Young}, Andr{\'e} and {Young}, Ken and {Younsi}, Ziri and {Yuan}, Feng and {Yuan}, Ye-Fei and {Zensus}, J. Anton and {Zhang}, Shuo and {Zhao}, Guang-Yao and {Zhao}, Shan-Shan and {Agurto}, Claudio and {Allardi}, Alexander and {Amestica}, Rodrigo and {Araneda}, Juan Pablo and {Arriagada}, Oriel and {Berghuis}, Jennie L. and {Bertarini}, Alessandra and {Berthold}, Ryan and {Blanchard}, Jay and {Brown}, Ken and {C{\'a}rdenas}, Mauricio and {Cantzler}, Michael and {Caro}, Patricio and {Castillo-Dom{\'\i}nguez}, Edgar and {Chan}, Tin Lok and {Chang}, Chih-Cheng and {Chang}, Dominic O. and {Chang}, Shu-Hao and {Chang}, Song-Chu and {Chen}, Chung-Chen and {Chilson}, Ryan and {Chuter}, Tim C. and {Ciechanowicz}, Miroslaw and {Colin-Beltran}, Edgar and {Coulson}, Iain M. and {Crowley}, Joseph and {Degenaar}, Nathalie and {Dornbusch}, Sven and {Dur{\'a}n}, Carlos A. and {Everett}, Wendeline B. and {Faber}, Aaron and {Forster}, Karl and {Fuchs}, Miriam M. and {Gale}, David M. and {Geertsema}, Gertie and {Gonz{\'a}lez}, Edouard and {Graham}, Dave and {Gueth}, Fr{\'e}d{\'e}ric and {Halverson}, Nils W. and {Han}, Chih-Chiang and {Han}, Kuo-Chang and {Hasegawa}, Yutaka and {Hern{\'a}ndez-Rebollar}, Jos{\'e} Luis and {Herrera}, Cristian and {Herrero-Illana}, Ruben and {Heyminck}, Stefan and {Hirota}, Akihiko and {Hoge}, James and {Hostler Schimpf}, Shelbi R. and {Howie}, Ryan E. and {Huang}, Yau-De and {Jiang}, Homin and {Jinchi}, Hao and {John}, David and {Kimura}, Kimihiro and {Klein}, Thomas and {Kubo}, Derek and {Kuroda}, John and {Kwon}, Caleb and {Lacasse}, Richard and {Laing}, Robert and {Leitch}, Erik M. and {Li}, Chao-Te and {Liu}, Ching-Tang and {Liu}, Kuan-Yu and {Lin}, Lupin C. -C. and {Lu}, Li-Ming and {Mac-Auliffe}, Felipe and {Martin-Cocher}, Pierre and {Matulonis}, Callie and {Maute}, John K. and {Messias}, Hugo and {Meyer-Zhao}, Zheng and {Monta{\~n}a}, Alfredo and {Montenegro-Montes}, Francisco and {Montgomerie}, William and {Moreno Nolasco}, Marcos Emir and {Muders}, Dirk and {Nishioka}, Hiroaki and {Norton}, Timothy J. and {Nystrom}, George and {Ogawa}, Hideo and {Olivares}, Rodrigo and {Oshiro}, Peter and {P{\'e}rez-Beaupuits}, Juan Pablo and {Parra}, Rodrigo and {Phillips}, Neil M. and {Poirier}, Michael and {Pradel}, Nicolas and {Qiu}, Richard and {Raffin}, Philippe A. and {Rahlin}, Alexandra S. and {Ram{\'\i}rez}, Jorge and {Ressler}, Sean and {Reynolds}, Mark and {Rodr{\'\i}guez-Montoya}, Iv{\'a}n and {Saez-Madain}, Alejandro F. and {Santana}, Jorge and {Shaw}, Paul and {Shirkey}, Leslie E. and {Silva}, Kevin M. and {Snow}, William and {Sousa}, Don and {Sridharan}, T.~K. and {Stahm}, William and {Stark}, Anthony A. and {Test}, John and {Torstensson}, Karl and {Venegas}, Paulina and {Walther}, Craig and {Wei}, Ta-Shun and {White}, Chris and {Wieching}, Gundolf and {Wijnands}, Rudy and {Wouterloot}, Jan G.~A. and {Yu}, Chen-Yu and {Yu (于威)}, Wei and {Zeballos}, Milagros},
	doi = {10.3847/2041-8213/ac667410.3847/2041-8213/ac667510.3847/2041-8213/ac6429},
	eid = {L12},
	journal = {\apjl},
	month = may,
	number = {2},
	pages = {L12},
	title = {{First Sagittarius A* Event Horizon Telescope Results. I. The Shadow of the Supermassive Black Hole in the Center of the Milky Way}},
	volume = {930},
	year = 2022}

@article{Swope1982,
	adsurl = {https://ui.adsabs.harvard.edu/abs/1982JChPh..76..637S},
	author = {{Swope}, William C. and {Andersen}, Hans C. and {Berens}, Peter H. and {Wilson}, Kent R.},
	doi = {10.1063/1.442716},
	journal = {The Journal of Chemical Physics},
	month = jan,
	number = {1},
	pages = {637-649},
	title = {{A computer simulation method for the calculation of equilibrium constants for the formation of physical clusters of molecules: Application to small water clusters}},
	volume = {76},
	year = 1982}

@article{Palumbo_2022,
	author = {Palumbo, Daniel C. M. and Gelles, Zachary and Tiede, Paul and Chang, Dominic O. and Pesce, Dominic W. and Chael, Andrew and Johnson, Michael D.},
	doi = {10.3847/1538-4357/ac9ab7},
	issn = {0004-637X},
	journal = {The Astrophysical Journal},
	langid = {english},
	month = nov,
	number = {2},
	pages = {107},
	publisher = {The American Astronomical Society},
	shorttitle = {Bayesian {{Accretion Modeling}}},
	title = {Bayesian {{Accretion Modeling}}: {{Axisymmetric Equatorial Emission}} in the {{Kerr Spacetime}}},
	volume = {939},
	year = {2022}}

@article{Tiede_2022,
	author = {Tiede, Paul},
	doi = {10.21105/joss.04457},
	journal = {Journal of Open Source Software},
	number = {76},
	pages = {4457},
	publisher = {The Open Journal},
	title = {Comrade: Composable Modeling of Radio Emission},
	url = {https://doi.org/10.21105/joss.04457},
	volume = {7},
	year = {2022}}

@article{Moscibrodzka_2023,
	author = {Moscibrodzka, Monika A. and Yfantis, Aristomenis I.},
	doi = {10.3847/1538-4365/acb6f9},
	journal = {The Astrophysical Journal Supplement Series},
	month = {mar},
	number = {1},
	pages = {22},
	publisher = {The American Astronomical Society},
	title = {Prospects for Ray-tracing Light Intensity and Polarization in Models of Accreting Compact Objects Using a GPU},
	url = {https://doi.org/10.3847/1538-4365/acb6f9},
	volume = {265},
	year = {2023}}

@article{Naethe_Motta_2025,
	author = {Naethe Motta, Pedro and Prather, Ben S. and C{\'a}rdenas-Avenda{\~n}o, Alejandro},
	doi = {10.3847/1538-4357/ae16a0},
	journal = {The Astrophysical Journal},
	month = {dec},
	number = {1},
	pages = {56},
	publisher = {The American Astronomical Society},
	title = {Jipole: A Differentiable ipole-based Code for Radiative Transfer in Curved Spacetimes},
	url = {https://doi.org/10.3847/1538-4357/ae16a0},
	volume = {995},
	year = {2025}}

@article{EHTM87Paper7,
	author = {{Event Horizon Telescope Collaboration} and Akiyama, Kazunori and Algaba, Juan Carlos and Alberdi, Antxon and Alef, Walter and Anantua, Richard and Asada, Keiichi and Azulay, Rebecca and Baczko, Anne-Kathrin and Ball, David and Balokovi{\'c}, Mislav and Barrett, John and Benson, Bradford A. and Bintley, Dan and Blackburn, Lindy and Blundell, Raymond and Boland, Wilfred and Bouman, Katherine L. and Bower, Geoffrey C. and Boyce, Hope and Bremer, Michael and Brinkerink, Christiaan D. and Brissenden, Roger and Britzen, Silke and Broderick, Avery E. and Broguiere, Dominique and Bronzwaer, Thomas and Byun, Do-Young and Carlstrom, John E. and Chael, Andrew and Chan, Chi-kwan and Chatterjee, Shami and Chatterjee, Koushik and Chen, Ming-Tang and Chen, Yongjun and Chesler, Paul M. and Cho, Ilje and Christian, Pierre and Conway, John E. and Cordes, James M. and Crawford, Thomas M. and Crew, Geoffrey B. and Cruz-Osorio, Alejandro and Cui, Yuzhu and Davelaar, Jordy and De Laurentis, Mariafelicia and Deane, Roger and Dempsey, Jessica and Desvignes, Gregory and Dexter, Jason and Doeleman, Sheperd S. and Eatough, Ralph P. and Falcke, Heino and Farah, Joseph and Fish, Vincent L. and Fomalont, Ed and Ford, H. Alyson and Fraga-Encinas, Raquel and Freeman, William T. and Friberg, Per and Fromm, Christian M. and Fuentes, Antonio and Galison, Peter and Gammie, Charles F. and Garc{\'\i}a, Roberto and Gentaz, Olivier and Georgiev, Boris and Goddi, Ciriaco and Gold, Roman and G{\'o}mez, Jos{\'e} L. and G{\'o}mez-Ruiz, Arturo I. and Gu, Minfeng and Gurwell, Mark and Hada, Kazuhiro and Haggard, Daryl and Hecht, Michael H. and Hesper, Ronald and Ho, Luis C. and Ho, Paul and Honma, Mareki and Huang, Chih-Wei L. and Huang, Lei and Hughes, David H. and Ikeda, Shiro and Inoue, Makoto and Issaoun, Sara and James, David J. and Jannuzi, Buell T. and Janssen, Michael and Jeter, Britton and Jiang, Wu and Jimenez-Rosales, Alejandra and Johnson, Michael D. and Jorstad, Svetlana and Jung, Taehyun and Karami, Mansour and Karuppusamy, Ramesh and Kawashima, Tomohisa and Keating, Garrett K. and Kettenis, Mark and Kim, Dong-Jin and Kim, Jae-Young and Kim, Jongsoo and Kim, Junhan and Kino, Motoki and Koay, Jun Yi and Kofuji, Yutaro and Koch, Patrick M. and Koyama, Shoko and Kramer, Michael and Kramer, Carsten and Krichbaum, Thomas P. and Kuo, Cheng-Yu and Lauer, Tod R. and Lee, Sang-Sung and Levis, Aviad and Li, Yan-Rong and Li, Zhiyuan and Lindqvist, Michael and Lico, Rocco and Lindahl, Greg and Liu, Jun and Liu, Kuo and Liuzzo, Elisabetta and Lo, Wen-Ping and Lobanov, Andrei P. and Loinard, Laurent and Lonsdale, Colin and Lu, Ru-Sen and MacDonald, Nicholas R. and Mao, Jirong and Marchili, Nicola and Markoff, Sera and Marrone, Daniel P. and Marscher, Alan P. and Mart{\'\i}-Vidal, Iv{\'a}n and Matsushita, Satoki and Matthews, Lynn D. and Medeiros, Lia and Menten, Karl M. and Mizuno, Izumi and Mizuno, Yosuke and Moran, James M. and Moriyama, Kotaro and Moscibrodzka, Monika and M{\"u}ller, Cornelia and Musoke, Gibwa and Mej{\'\i}as, Alejandro Mus and Michalik, Daniel and Nadolski, Andrew and Nagai, Hiroshi and Nagar, Neil M. and Nakamura, Masanori and Narayan, Ramesh and Narayanan, Gopal and Natarajan, Iniyan and Nathanail, Antonios and Neilsen, Joey and Neri, Roberto and Ni, Chunchong and Noutsos, Aristeidis and Nowak, Michael A. and Okino, Hiroki and Olivares, H{\'e}ctor and Ortiz-Le{\'o}n, Gisela N. and Oyama, Tomoaki and {\"O}zel, Feryal and Palumbo, Daniel C. M. and Park, Jongho and Patel, Nimesh and Pen, Ue-Li and Pesce, Dominic W. and Pi{\'e}tu, Vincent and Plambeck, Richard and PopStefanija, Aleksandar and Porth, Oliver and P{\"o}tzl, Felix M. and Prather, Ben and Preciado-L{\'o}pez, Jorge A. and Psaltis, Dimitrios and Pu, Hung-Yi and Ramakrishnan, Venkatessh and Rao, Ramprasad and Rawlings, Mark G. and Raymond, Alexander W. and Rezzolla, Luciano and Ricarte, Angelo and Ripperda, Bart and Roelofs, Freek and Rogers, Alan and Ros, Eduardo and Rose, Mel and Roshanineshat, Arash and Rottmann, Helge and Roy, Alan L. and Ruszczyk, Chet and Rygl, Kazi L. J. and S{\'a}nchez, Salvador and S{\'a}nchez-Arguelles, David and Sasada, Mahito and Savolainen, Tuomas and Schloerb, F. Peter and Schuster, Karl-Friedrich and Shao, Lijing and Shen, Zhiqiang and Small, Des and Sohn, Bong Won and SooHoo, Jason and Sun, He and Tazaki, Fumie and Tetarenko, Alexandra J. and Tiede, Paul and Tilanus, Remo P. J. and Titus, Michael and Toma, Kenji and Torne, Pablo and Trent, Tyler and Traianou, Efthalia and Trippe, Sascha and Bemmel, Ilse van and van Langevelde, Huib Jan and van Rossum, Daniel R. and Wagner, Jan and Ward-Thompson, Derek and Wardle, John and Weintroub, Jonathan and Wex, Norbert and Wharton, Robert and Wielgus, Maciek and Wong, George N. and Wu, Qingwen and Yoon, Doosoo and Young, Andr{\'e} and Young, Ken and Younsi, Ziri and Yuan, Feng and Yuan, Ye-Fei and Zensus, J. Anton and Zhao, Guang-Yao and Zhao, Shan-Shan and The Event Horizon Telescope Collaboration},
	doi = {10.3847/2041-8213/abe71d},
	journal = {The Astrophysical Journal Letters},
	month = {mar},
	number = {1},
	pages = {L12},
	publisher = {The American Astronomical Society},
	title = {First M87 Event Horizon Telescope Results. VII. Polarization of the Ring},
	url = {https://doi.org/10.3847/2041-8213/abe71d},
	volume = {910},
	year = {2021}}

@article{Prather_2021,
	adsurl = {https://ui.adsabs.harvard.edu/abs/2021JOSS....6.3336P},
	archiveprefix = {arXiv},
	author = {{Prather}, Ben and {Wong}, George and {Dhruv}, Vedant and {Ryan}, Benjamin and {Dolence}, Joshua and {Ressler}, Sean and {Gammie}, Charles},
	doi = {10.21105/joss.03336},
	eid = {3336},
	eprint = {2110.10191},
	journal = {The Journal of Open Source Software},
	month = oct,
	number = {66},
	pages = {3336},
	primaryclass = {astro-ph.HE},
	title = {{iharm3D: Vectorized General Relativistic Magnetohydrodynamics}},
	volume = {6},
	year = 2021}

@article{Prather_2023,
	author = {Prather, Ben S. and Dexter, Jason and Moscibrodzka, Monika and Pu, Hung-Yi and Bronzwaer, Thomas and Davelaar, Jordy and Younsi, Ziri and Gammie, Charles F. and Gold, Roman and Wong, George N. and Akiyama, Kazunori and Alberdi, Antxon and Alef, Walter and Algaba, Juan Carlos and Anantua, Richard and Asada, Keiichi and Azulay, Rebecca and Bach, Uwe and Baczko, Anne-Kathrin and Ball, David and Balokovi{\'c}, Mislav and Barrett, John and Baub{\"o}ck, Michi and Benson, Bradford A. and Bintley, Dan and Blackburn, Lindy and Blundell, Raymond and Bouman, Katherine L. and Bower, Geoffrey C. and Boyce, Hope and Bremer, Michael and Brinkerink, Christiaan D. and Brissenden, Roger and Britzen, Silke and Broderick, Avery E. and Broguiere, Dominique and Bustamante, Sandra and Byun, Do-Young and Carlstrom, John E. and Ceccobello, Chiara and Chael, Andrew and Chan, Chi-kwan and Chang, Dominic O. and Chatterjee, Koushik and Chatterjee, Shami and Chen, Ming-Tang and Chen, Yongjun and Cheng, Xiaopeng and Cho, Ilje and Christian, Pierre and Conroy, Nicholas S. and Conway, John E. and Cordes, James M. and Crawford, Thomas M. and Crew, Geoffrey B. and Cruz-Osorio, Alejandro and Cui, Yuzhu and De Laurentis, Mariafelicia and Deane, Roger and Dempsey, Jessica and Desvignes, Gregory and Dhruv, Vedant and Doeleman, Sheperd S. and Dougal, Sean and Dzib, Sergio A. and Eatough, Ralph P. and Emami, Razieh and Falcke, Heino and Farah, Joseph and Fish, Vincent L. and Fomalont, Ed and Ford, H. Alyson and Fraga-Encinas, Raquel and Freeman, William T. and Friberg, Per and Fromm, Christian M. and Fuentes, Antonio and Galison, Peter and Garc{\'\i}a, Roberto and Gentaz, Olivier and Georgiev, Boris and Goddi, Ciriaco and G{\'o}mez-Ruiz, Arturo I. and G{\'o}mez, Jos{\'e} L. and Gu, Minfeng and Gurwell, Mark and Hada, Kazuhiro and Haggard, Daryl and Haworth, Kari and Hecht, Michael H. and Hesper, Ronald and Heumann, Dirk and Ho, Luis C. and Ho, Paul and Honma, Mareki and Huang, Chih-Wei L. and Huang, Lei and Hughes, David H. and Ikeda, Shiro and Impellizzeri, C. M. Violette and Inoue, Makoto and Issaoun, Sara and James, David J. and Jannuzi, Buell T. and Janssen, Michael and Jeter, Britton and Jiang, Wu and Jim{\'e}nez-Rosales, Alejandra and Johnson, Michael D. and Jorstad, Svetlana and Joshi, Abhishek V. and Jung, Taehyun and Karami, Mansour and Karuppusamy, Ramesh and Kawashima, Tomohisa and Keating, Garrett K. and Kettenis, Mark and Kim, Dong-Jin and Kim, Jae-Young and Kim, Jongsoo and Kim, Junhan and Kino, Motoki and Koay, Jun Yi and Kocherlakota, Prashant and Kofuji, Yutaro and Koyama, Shoko and Kramer, Carsten and Kramer, Michael and Krichbaum, Thomas P. and Kuo, Cheng-Yu and La Bella, Noemi and Lauer, Tod R. and Lee, Daeyoung and Lee, Sang-Sung and Leung, Po Kin and Levis, Aviad and Li, Zhiyuan and Lico, Rocco and Lindahl, Greg and Lindqvist, Michael and Lisakov, Mikhail and Liu, Jun and Liu, Kuo and Liuzzo, Elisabetta and Lo, Wen-Ping and Lobanov, Andrei P. and Loinard, Laurent and Lonsdale, Colin J. and Lu, Ru-Sen and MacDonald, Nicholas R. and Mao, Jirong and Marchili, Nicola and Markoff, Sera and Marrone, Daniel P. and Marscher, Alan P. and Mart{\'\i}-Vidal, Iv{\'a}n and Matsushita, Satoki and Matthews, Lynn D. and Medeiros, Lia and Menten, Karl M. and Michalik, Daniel and Mizuno, Izumi and Mizuno, Yosuke and Moran, James M. and Moriyama, Kotaro and M{\"u}ller, Cornelia and Mus, Alejandro and Musoke, Gibwa and Myserlis, Ioannis and Nadolski, Andrew and Nagai, Hiroshi and Nagar, Neil M. and Nakamura, Masanori and Narayan, Ramesh and Narayanan, Gopal and Natarajan, Iniyan and Nathanail, Antonios and Fuentes, Santiago Navarro and Neilsen, Joey and Neri, Roberto and Ni, Chunchong and Noutsos, Aristeidis and Nowak, Michael A. and Oh, Junghwan and Okino, Hiroki and Olivares, H{\'e}ctor and Ortiz-Le{\'o}n, Gisela N. and Oyama, Tomoaki and {\"O}zel, Feryal and Palumbo, Daniel C. M. and Paraschos, Georgios Filippos and Park, Jongho and Parsons, Harriet and Patel, Nimesh and Pen, Ue-Li and Pesce, Dominic W. and Pi{\'e}tu, Vincent and Plambeck, Richard and PopStefanija, Aleksandar and Porth, Oliver and P{\"o}tzl, Felix M. and Preciado-L{\'o}pez, Jorge A. and Psaltis, Dimitrios and Ramakrishnan, Venkatessh and Rao, Ramprasad and Rawlings, Mark G. and Raymond, Alexander W. and Rezzolla, Luciano and Ricarte, Angelo and Ripperda, Bart and Roelofs, Freek and Rogers, Alan and Ros, Eduardo and Romero-Ca{\~n}izales, Cristina and Roshanineshat, Arash and Rottmann, Helge and Roy, Alan L. and Ruiz, Ignacio and Ruszczyk, Chet and Rygl, Kazi L. J. and S{\'a}nchez, Salvador and S{\'a}nchez-Arg{\"u}elles, David and S{\'a}nchez-Portal, Miguel and Sasada, Mahito and Satapathy, Kaushik and Savolainen, Tuomas and Schloerb, F. Peter and Schonfeld, Jonathan and Schuster, Karl-Friedrich and Shao, Lijing and Shen, Zhiqiang and Small, Des and Sohn, Bong Won and SooHoo, Jason and Souccar, Kamal and Sun, He and Tazaki, Fumie and Tetarenko, Alexandra J. and Tiede, Paul and Tilanus, Remo P. J. and Titus, Michael and Torne, Pablo and Traianou, Efthalia and Trent, Tyler and Trippe, Sascha and Turk, Matthew and van Bemmel, Ilse and van Langevelde, Huib Jan and van Rossum, Daniel R. and Vos, Jesse and Wagner, Jan and Ward-Thompson, Derek and Wardle, John and Weintroub, Jonathan and Wex, Norbert and Wharton, Robert and Wielgus, Maciek and Wiik, Kaj and Witzel, Gunther and Wondrak, Michael F. and Wu, Qingwen and Yamaguchi, Paul and Yfantis, Aristomenis and Yoon, Doosoo and Young, Andr{\'e} and Young, Ken and Yu, Wei and Yuan, Feng and Yuan, Ye-Fei and Zensus, J. Anton and Zhang, Shuo and Zhao, Guang-Yao and Zhao, Shan-Shan and The Event Horizon Telescope Collaboration},
	doi = {10.3847/1538-4357/acc586},
	journal = {The Astrophysical Journal},
	month = {jun},
	number = {1},
	pages = {35},
	publisher = {The American Astronomical Society},
	title = {Comparison of Polarized Radiative Transfer Codes Used by the EHT Collaboration},
	url = {https://doi.org/10.3847/1538-4357/acc586},
	volume = {950},
	year = {2023}}

@article{Johnson:2023ynn,
	adsurl = {https://ui.adsabs.harvard.edu/abs/2023Galax..11...61J},
	archiveprefix = {arXiv},
	author = {{Johnson}, Michael D. and {Akiyama}, Kazunori and {Blackburn}, Lindy and {Bouman}, Katherine L. and {Broderick}, Avery E. and {Cardoso}, Vitor and {Fender}, Rob P. and {Fromm}, Christian M. and {Galison}, Peter and {G{\'o}mez}, Jos{\'e} L. and {Haggard}, Daryl and {Lister}, Matthew L. and {Lobanov}, Andrei P. and {Markoff}, Sera and {Narayan}, Ramesh and {Natarajan}, Priyamvada and {Nichols}, Tiffany and {Pesce}, Dominic W. and {Younsi}, Ziri and {Chael}, Andrew and {Chatterjee}, Koushik and {Chaves}, Ryan and {Doboszewski}, Juliusz and {Dodson}, Richard and {Doeleman}, Sheperd S. and {Elder}, Jamee and {Fitzpatrick}, Garret and {Haworth}, Kari and {Houston}, Janice and {Issaoun}, Sara and {Kovalev}, Yuri Y. and {Levis}, Aviad and {Lico}, Rocco and {Marcoci}, Alexandru and {Martens}, Niels C.~M. and {Nagar}, Neil M. and {Oppenheimer}, Aaron and {Palumbo}, Daniel C.~M. and {Ricarte}, Angelo and {Rioja}, Mar{\'\i}a J. and {Roelofs}, Freek and {Thresher}, Ann C. and {Tiede}, Paul and {Weintroub}, Jonathan and {Wielgus}, Maciek},
	doi = {10.3390/galaxies11030061},
	eid = {61},
	eprint = {2304.11188},
	journal = {Galaxies},
	month = apr,
	number = {3},
	pages = {61},
	primaryclass = {astro-ph.HE},
	title = {{Key Science Goals for the Next-Generation Event Horizon Telescope}},
	volume = {11},
	year = 2023}

@inproceedings{Johnson:2024ttr,
	adsurl = {https://ui.adsabs.harvard.edu/abs/2024SPIE13092E..2DJ},
	archiveprefix = {arXiv},
	author = {{Johnson}, Michael D. and {Akiyama}, Kazunori and {Baturin}, Rebecca and {Bilyeu}, Bryan and {Blackburn}, Lindy and {Boroson}, Don and {C{\'a}rdenas-Avenda{\~n}o}, Alejandro and {Chael}, Andrew and {Chan}, Chi-kwan and {Chang}, Dominic and {Cheimets}, Peter and {Chou}, Cathy and {Doeleman}, Sheperd S. and {Farah}, Joseph and {Galison}, Peter and {Gamble}, Ronald and {Gammie}, Charles F. and {Gelles}, Zachary and {G{\'o}mez}, Jos{\'e} L. and {Gralla}, Samuel E. and {Grimes}, Paul and {Gurvits}, Leonid I. and {Hadar}, Shahar and {Haworth}, Kari and {Hada}, Kazuhiro and {Hecht}, Michael H. and {Honma}, Mareki and {Houston}, Janice and {Hudson}, Ben and {Issaoun}, Sara and {Jia}, He and {Jorstad}, Svetlana and {Kauffman}, Jens and {Kovalev}, Yuri Y. and {Kurczynski}, Peter and {Lafon}, Robert E. and {Lupsasca}, Alexandru and {Lehmensiek}, Robert and {Ma}, Chung-Pei and {Marrone}, Daniel P. and {Marscher}, Alan P. and {Melnick}, Gary and {Narayan}, Ramesh and {Niinuma}, Kotaro and {Noble}, Scott C. and {Palmer}, Eric J. and {Palumbo}, Daniel C.~M. and {Paritsky}, Lenny and {Peretz}, Eliad and {Pesce}, Dominic and {Plavin}, Alexander and {Quataert}, Eliot and {Rana}, Hannah and {Ricarte}, Angelo and {Roelofs}, Freek and {Shtyrkova}, Katia and {Sinclair}, Laura C. and {Small}, Jeffrey and {Kumara}, Sridharan Tirupati and {Srinivasan}, Ranjani and {Strominger}, Andrew and {Tiede}, Paul and {Tong}, Edward and {Wang}, Jade and {Weintroub}, Jonathan and {Wielgus}, Maciek and {Wong}, George},
	booktitle = {Space Telescopes and Instrumentation 2024: Optical, Infrared, and Millimeter Wave},
	doi = {10.1117/12.3019835},
	editor = {{Coyle}, Laura E. and {Matsuura}, Shuji and {Perrin}, Marshall D.},
	eid = {130922D},
	eprint = {2406.12917},
	month = aug,
	pages = {130922D},
	primaryclass = {astro-ph.IM},
	series = {Society of Photo-Optical Instrumentation Engineers (SPIE) Conference Series},
	title = {{The Black Hole Explorer: motivation and vision}},
	volume = {13092},
	year = 2024}

@article{Monika_2018,
	adsurl = {https://ui.adsabs.harvard.edu/abs/2018MNRAS.475...43M},
	archiveprefix = {arXiv},
	author = {{Mo{\'s}cibrodzka}, M. and {Gammie}, C.~F.},
	doi = {10.1093/mnras/stx3162},
	eprint = {1712.03057},
	journal = {\mnras},
	month = mar,
	number = {1},
	pages = {43-54},
	primaryclass = {astro-ph.HE},
	title = {{IPOLE - semi-analytic scheme for relativistic polarized radiative transport}},
	volume = {475},
	year = 2018}

@article{EHTSgrAPaper5,
	author = {{Event Horizon Telescope Collaboration} and Akiyama, Kazunori and Alberdi, Antxon and Alef, Walter and Algaba, Juan Carlos and Anantua, Richard and Asada, Keiichi and Azulay, Rebecca and Bach, Uwe and Baczko, Anne-Kathrin and Ball, David and Balokovi\'c, Mislav and Barrett, John and Baub\"ock, Michi and Benson, Bradford A. and Bintley, Dan and Blackburn, Lindy and Blundell, Raymond and Bouman, Katherine L. and Bower, Geoffrey C. and Boyce, Hope and Bremer, Michael and Brinkerink, Christiaan D. and Brissenden, Roger and Britzen, Silke and Broderick, Avery E. and Broguiere, Dominique and Bronzwaer, Thomas and Bustamante, Sandra and Byun, Do-Young and Carlstrom, John E. and Ceccobello, Chiara and Chael, Andrew and Chan, Chi-kwan and Chatterjee, Koushik and Chatterjee, Shami and Chen, Ming-Tang and Chen, Yongjun and Cheng, Xiaopeng and Cho, Ilje and Christian, Pierre and Conroy, Nicholas S. and Conway, John E. and Cordes, James M. and Crawford, Thomas M. and Crew, Geoffrey B. and Cruz-Osorio, Alejandro and Cui, Yuzhu and Davelaar, Jordy and Laurentis, Mariafelicia De and Deane, Roger and Dempsey, Jessica and Desvignes, Gregory and Dexter, Jason and Dhruv, Vedant and Doeleman, Sheperd S. and Dougal, Sean and Dzib, Sergio A. and Eatough, Ralph P. and Emami, Razieh and Falcke, Heino and Farah, Joseph and Fish, Vincent L. and Fomalont, Ed and Ford, H. Alyson and Fraga-Encinas, Raquel and Freeman, William T. and Friberg, Per and Fromm, Christian M. and Fuentes, Antonio and Galison, Peter and Gammie, Charles F. and Garc\'ia, Roberto and Gentaz, Olivier and Georgiev, Boris and Goddi, Ciriaco and Gold, Roman and G\'omez-Ruiz, Arturo I. and G\'omez, Jos\'e L. and Gu, Minfeng and Gurwell, Mark and Hada, Kazuhiro and Haggard, Daryl and Haworth, Kari and Hecht, Michael H. and Hesper, Ronald and Heumann, Dirk and Ho, Luis C. and Ho, Paul and Honma, Mareki and Huang, Chih-Wei L. and Huang, Lei and Hughes, David H. and Ikeda, Shiro and Impellizzeri, C. M. Violette and Inoue, Makoto and Issaoun, Sara and James, David J. and Jannuzi, Buell T. and Janssen, Michael and Jeter, Britton and Jiang, Wu and Jim\'enez-Rosales, Alejandra and Johnson, Michael D. and Jorstad, Svetlana and Joshi, Abhishek V. and Jung, Taehyun and Karami, Mansour and Karuppusamy, Ramesh and Kawashima, Tomohisa and Keating, Garrett K. and Kettenis, Mark and Kim, Dong-Jin and Kim, Jae-Young and Kim, Jongsoo and Kim, Junhan and Kino, Motoki and Koay, Jun Yi and Kocherlakota, Prashant and Kofuji, Yutaro and Koch, Patrick M. and Koyama, Shoko and Kramer, Carsten and Kramer, Michael and Krichbaum, Thomas P. and Kuo, Cheng-Yu and Bella, Noemi La and Lauer, Tod R. and Lee, Daeyoung and Lee, Sang-Sung and Leung, Po Kin and Levis, Aviad and Li, Zhiyuan and Lico, Rocco and Lindahl, Greg and Lindqvist, Michael and Lisakov, Mikhail and Liu, Jun and Liu, Kuo and Liuzzo, Elisabetta and Lo, Wen-Ping and Lobanov, Andrei P. and Loinard, Laurent and Lonsdale, Colin J. and Lu, Ru-Sen and Mao, Jirong and Marchili, Nicola and Markoff, Sera and Marrone, Daniel P. and Marscher, Alan P. and Mart\'i-Vidal, Iv\'an and Matsushita, Satoki and Matthews, Lynn D. and Medeiros, Lia and Menten, Karl M. and Michalik, Daniel and Mizuno, Izumi and Mizuno, Yosuke and Moran, James M. and Moriyama, Kotaro and Moscibrodzka, Monika and M\"uller, Cornelia and Mus, Alejandro and Musoke, Gibwa and Myserlis, Ioannis and Nadolski, Andrew and Nagai, Hiroshi and Nagar, Neil M. and Nakamura, Masanori and Narayan, Ramesh and Narayanan, Gopal and Natarajan, Iniyan and Nathanail, Antonios and Fuentes, Santiago Navarro and Neilsen, Joey and Neri, Roberto and Ni, Chunchong and Noutsos, Aristeidis and Nowak, Michael A. and Oh, Junghwan and Okino, Hiroki and Olivares, H\'ector and Ortiz-Le\'on, Gisela N. and Oyama, Tomoaki and \"Ozel, Feryal and Palumbo, Daniel C. M. and Paraschos, Georgios Filippos and Park, Jongho and Parsons, Harriet and Patel, Nimesh and Pen, Ue-Li and Pesce, Dominic W. and Pi\'etu, Vincent and Plambeck, Richard and PopStefanija, Aleksandar and Porth, Oliver and P\"otzl, Felix M. and Prather, Ben and Preciado-L\'opez, Jorge A. and Psaltis, Dimitrios and Pu, Hung-Yi and Ramakrishnan, Venkatessh and Rao, Ramprasad and Rawlings, Mark G. and Raymond, Alexander W. and Rezzolla, Luciano and Ricarte, Angelo and Ripperda, Bart and Roelofs, Freek and Rogers, Alan and Ros, Eduardo and Romero-Ca\~nizales, Cristina and Roshanineshat, Arash and Rottmann, Helge and Roy, Alan L. and Ruiz, Ignacio and Ruszczyk, Chet and Rygl, Kazi L. J. and S\'anchez, Salvador and S\'anchez-Arg\"uelles, David and S\'anchez-Portal, Miguel and Sasada, Mahito and Satapathy, Kaushik and Savolainen, Tuomas and Schloerb, F. Peter and Schonfeld, Jonathan and Schuster, Karl-Friedrich and Shao, Lijing and Shen, Zhiqiang and Small, Des and Sohn, Bong Won and SooHoo, Jason and Souccar, Kamal and Sun, He and Tazaki, Fumie and Tetarenko, Alexandra J. and Tiede, Paul and Tilanus, Remo P. J. and Titus, Michael and Torne, Pablo and Traianou, Efthalia and Trent, Tyler and Trippe, Sascha and Turk, Matthew and van Bemmel, Ilse and van Langevelde, Huib Jan and van Rossum, Daniel R. and Vos, Jesse and Wagner, Jan and Ward-Thompson, Derek and Wardle, John and Weintroub, Jonathan and Wex, Norbert and Wharton, Robert and Wielgus, Maciek and Wiik, Kaj and Witzel, Gunther and Wondrak, Michael F. and Wong, George N. and Wu, Qingwen and Yamaguchi, Paul and Yoon, Doosoo and Young, Andr\'e and Young, Ken and Younsi, Ziri and Yuan, Feng and Yuan, Ye-Fei and Zensus, J. Anton and Zhang, Shuo and Zhao, Guang-Yao and Zhao, Shan-Shan and Chan, Tin Lok and Qiu, Richard and Ressler, Sean and White, Chris},
	date = {2022-05-12},
	doi = {10.3847/2041-8213/ac6672},
	issn = {2041-8205},
	journal = {The Astrophysical Journal Letters},
	langid = {english},
	number = {2},
	pages = {L16},
	publisher = {IOP Publishing},
	shortjournal = {ApJL},
	title = {First {{Sagittarius A}}* {{Event Horizon Telescope Results}}. {{V}}. {{Testing Astrophysical Models}} of the {{Galactic Center Black Hole}}},
	volume = {930},
	year = {2022}}

@article{Wong:2022rqr,
	archiveprefix = {arXiv},
	author = {Wong, George N. and others},
	doi = {10.3847/1538-4365/ac582e},
	eprint = {2202.11721},
	journal = {Astrophys. J. Supp.},
	number = {2},
	pages = {64},
	primaryclass = {astro-ph.HE},
	title = {{PATOKA: Simulating Electromagnetic Observables of Black Hole Accretion}},
	volume = {259},
	year = {2022}}

@article{EHTM87Paper5,
	author = {{Event Horizon Telescope Collaboration} and Akiyama, Kazunori and Alberdi, Antxon and Alef, Walter and Asada, Keiichi and Azulay, Rebecca and Baczko, Anne-Kathrin and Ball, David and Balokovi\'c, Mislav and Barrett, John and Bintley, Dan and Blackburn, Lindy and Boland, Wilfred and Bouman, Katherine L. and Bower, Geoffrey C. and Bremer, Michael and Brinkerink, Christiaan D. and Brissenden, Roger and Britzen, Silke and Broderick, Avery E. and Broguiere, Dominique and Bronzwaer, Thomas and Byun, Do-Young and Carlstrom, John E. and Chael, Andrew and Chan, Chi-kwan and Chatterjee, Shami and Chatterjee, Koushik and Chen, Ming-Tang and Chen, Yongjun and Cho, Ilje and Christian, Pierre and Conway, John E. and Cordes, James M. and Crew, Geoffrey B. and Cui, Yuzhu and Davelaar, Jordy and De Laurentis, Mariafelicia and Deane, Roger and Dempsey, Jessica and Desvignes, Gregory and Dexter, Jason and Doeleman, Sheperd S. and Eatough, Ralph P. and Falcke, Heino and Fish, Vincent L. and Fomalont, Ed and Fraga-Encinas, Raquel and Friberg, Per and Fromm, Christian M. and G\'omez, Jos\'e L. and Galison, Peter and Gammie, Charles F. and Garc\'ia, Roberto and Gentaz, Olivier and Georgiev, Boris and Goddi, Ciriaco and Gold, Roman and Gu, Minfeng and Gurwell, Mark and Hada, Kazuhiro and Hecht, Michael H. and Hesper, Ronald and Ho, Luis C. and Ho, Paul and Honma, Mareki and Huang, Chih-Wei L. and Huang, Lei and Hughes, David H. and Ikeda, Shiro and Inoue, Makoto and Issaoun, Sara and James, David J. and Jannuzi, Buell T. and Janssen, Michael and Jeter, Britton and Jiang, Wu and Johnson, Michael D. and Jorstad, Svetlana and Jung, Taehyun and Karami, Mansour and Karuppusamy, Ramesh and Kawashima, Tomohisa and Keating, Garrett K. and Kettenis, Mark and Kim, Jae-Young and Kim, Junhan and Kim, Jongsoo and Kino, Motoki and Koay, Jun Yi and Koch, Patrick M. and Koyama, Shoko and Kramer, Michael and Kramer, Carsten and Krichbaum, Thomas P. and Kuo, Cheng-Yu and Lauer, Tod R. and Lee, Sang-Sung and Li, Yan-Rong and Li, Zhiyuan and Lindqvist, Michael and Liu, Kuo and Liuzzo, Elisabetta and Lo, Wen-Ping and Lobanov, Andrei P. and Loinard, Laurent and Lonsdale, Colin and Lu, Ru-Sen and MacDonald, Nicholas R. and Mao, Jirong and Markoff, Sera and Marrone, Daniel P. and Marscher, Alan P. and Mart\'i-Vidal, Iv\'an and Matsushita, Satoki and Matthews, Lynn D. and Medeiros, Lia and Menten, Karl M. and Mizuno, Yosuke and Mizuno, Izumi and Moran, James M. and Moriyama, Kotaro and Moscibrodzka, Monika and Mu\"ller, Cornelia and Nagai, Hiroshi and Nagar, Neil M. and Nakamura, Masanori and Narayan, Ramesh and Narayanan, Gopal and Natarajan, Iniyan and Neri, Roberto and Ni, Chunchong and Noutsos, Aristeidis and Okino, Hiroki and Olivares, H\'ector and Oyama, Tomoaki and \"Ozel, Feryal and Palumbo, Daniel C. M. and Patel, Nimesh and Pen, Ue-Li and Pesce, Dominic W. and Pi\'etu, Vincent and Plambeck, Richard and PopStefanija, Aleksandar and Porth, Oliver and Prather, Ben and Preciado-L\'opez, Jorge A. and Psaltis, Dimitrios and Pu, Hung-Yi and Ramakrishnan, Venkatessh and Rao, Ramprasad and Rawlings, Mark G. and Raymond, Alexander W. and Rezzolla, Luciano and Ripperda, Bart and Roelofs, Freek and Rogers, Alan and Ros, Eduardo and Rose, Mel and Roshanineshat, Arash and Rottmann, Helge and Roy, Alan L. and Ruszczyk, Chet and Ryan, Benjamin R. and Rygl, Kazi L. J. and S\'anchez, Salvador and S\'anchez-Arguelles, David and Sasada, Mahito and Savolainen, Tuomas and Schloerb, F. Peter and Schuster, Karl-Friedrich and Shao, Lijing and Shen, Zhiqiang and Small, Des and Sohn, Bong Won and SooHoo, Jason and Tazaki, Fumie and Tiede, Paul and Tilanus, Remo P. J. and Titus, Michael and Toma, Kenji and Torne, Pablo and Trent, Tyler and Trippe, Sascha and Tsuda, Shuichiro and van Bemmel, Ilse and van Langevelde, Huib Jan and van Rossum, Daniel R. and Wagner, Jan and Wardle, John and Weintroub, Jonathan and Wex, Norbert and Wharton, Robert and Wielgus, Maciek and Wong, George N. and Wu, Qingwen and Young, Andr\'e and Young, Ken and Younsi, Ziri and Yuan, Feng and Yuan, Ye-Fei and Zensus, J. Anton and Zhao, Guangyao and Zhao, Shan-Shan and Zhu, Ziyan and Anczarski, Jadyn and Baganoff, Frederick K. and Eckart, Andreas and Farah, Joseph R. and Haggard, Daryl and Meyer-Zhao, Zheng and Michalik, Daniel and Nadolski, Andrew and Neilsen, Joseph and Nishioka, Hiroaki and Nowak, Michael A. and Pradel, Nicolas and Primiani, Rurik A. and Souccar, Kamal and Vertatschitsch, Laura and Yamaguchi, Paul and Zhang, Shuo},
	date = {2019-04-01},
	doi = {10.3847/2041-8213/ab0f43},
	issn = {0004-637X},
	journal = {The Astrophysical Journal},
	options = {useprefix=true},
	pages = {L5},
	title = {First {{M87 Event Horizon Telescope Results}}. {{V}}. {{Physical Origin}} of the {{Asymmetric Ring}}},
	volume = {875},
	year = {2019}}

@book{Mihalas_1984,
	adsurl = {https://ui.adsabs.harvard.edu/abs/1984oup..book.....M},
	author = {{Mihalas}, D. and {Mihalas}, B.~W.},
	title = {{Foundations of radiation hydrodynamics}},
	year = 1984}

@article{Sharma:2023nbk,
	archiveprefix = {arXiv},
	author = {Sharma, Aniket and Medeiros, Lia and Wong, George N. and Chan, Chi-kwan and Halevi, Goni and Mullen, Patrick D. and Stone, James M.},
	doi = {10.3847/1538-4357/adc104},
	eprint = {2304.03804},
	journal = {Astrophys. J.},
	number = {1},
	pages = {40},
	primaryclass = {astro-ph.HE},
	title = {{Mahakala: A Python-based Modular Ray-tracing and Radiative Transfer Algorithm for Curved Spacetimes}},
	volume = {985},
	year = {2025}}

@article{Chang_2024,
	author = {Chang, Dominic},
	doi = {10.21105/joss.07273},
	journal = {Journal of Open Source Software},
	number = {102},
	pages = {7273},
	publisher = {The Open Journal},
	title = {Krang: Kerr Raytracer for Analytic Null Geodesics},
	url = {https://doi.org/10.21105/joss.07273},
	volume = {9},
	year = {2024}}

@article{Keeble_2025,
	adsurl = {https://ui.adsabs.harvard.edu/abs/2025PhRvD.111j3042K},
	archiveprefix = {arXiv},
	author = {{Keeble}, Lennox S. and {C{\'a}rdenas-Avenda{\~n}o}, Alejandro and {Palumbo}, Daniel C.~M.},
	doi = {10.1103/PhysRevD.111.103042},
	eid = {103042},
	eprint = {2502.20312},
	journal = {\prd},
	month = may,
	number = {10},
	pages = {103042},
	primaryclass = {astro-ph.HE},
	title = {{Inferring black hole spin from interferometric measurements of the first photon ring: A geometric approach}},
	volume = {111},
	year = 2025}

@article{Canfield_1987,
	adsurl = {https://ui.adsabs.harvard.edu/abs/1987ApJ...323..565C},
	author = {{Canfield}, E. and {Howard}, W.~M. and {Liang}, E.~P.},
	doi = {10.1086/165853},
	journal = {\apj},
	month = dec,
	pages = {565-574},
	title = {{Inverse Comptonization by one-dimensional relativistic electrons}},
	volume = {323},
	year = 1987}

@article{Yfantis_2024_bipole,
	author = {Yfantis, A. I. and Mo{\'s}cibrodzka, M. A. and Wielgus, M. and Vos, J. T. and Jimenez-Rosales, A.},
	doi = {10.1051/0004-6361/202348230},
	issn = {1432-0746},
	journal = {Astronomy \&; Astrophysics},
	month = may,
	pages = {A142},
	publisher = {EDP Sciences},
	title = {Fitting the light curves of Sagittarius A* with a hot-spot model: Bayesian modeling of QU loops in the millimeter band},
	url = {http://dx.doi.org/10.1051/0004-6361/202348230},
	volume = {685},
	year = {2024}}

@book{Landau_1975,
	address = {Oxford},
	author = {Landau, L. D. and Lifschits, E. M.},
	isbn = {978-0-08-018176-9},
	publisher = {Pergamon Press},
	series = {Course of Theoretical Physics},
	title = {{The Classical Theory of Fields}},
	volume = {Volume 2},
	year = {1975}}

@article{kahn_1950,
	author = {Kahn, H.},
	journal = {Nucleonics},
	month = jun,
	number = {6},
	pages = {60--65},
	pmid = {15423762},
	title = {Random sampling (Monte Carlo) techniques in neutron attenuation problems--II},
	volume = {6},
	year = {1950}}

@article{Younsi_2012,
	author = {Younsi, Z. and Wu, K. and Fuerst, S. V.},
	doi = {10.1051/0004-6361/201219599},
	issn = {1432-0746},
	journal = {Astronomy \&; Astrophysics},
	month = aug,
	pages = {A13},
	publisher = {EDP Sciences},
	title = {General relativistic radiative transfer: formulation and emission from structured tori around black holes},
	url = {http://dx.doi.org/10.1051/0004-6361/201219599},
	volume = {545},
	year = {2012}}

@article{Dolence_2009,
	author = {Dolence, Joshua C. and Gammie, Charles F. and Mo{\'s}cibrodzka, Monika and Leung, Po Kin},
	doi = {10.1088/0067-0049/184/2/387},
	issn = {1538-4365},
	journal = {The Astrophysical Journal Supplement Series},
	month = oct,
	number = {2},
	pages = {387--397},
	publisher = {American Astronomical Society},
	title = {grmonty: A MONTE CARLO CODE FOR RELATIVISTIC RADIATIVE TRANSPORT},
	url = {http://dx.doi.org/10.1088/0067-0049/184/2/387},
	volume = {184},
	year = {2009}}

@article{Leung_2011,
	adsurl = {https://ui.adsabs.harvard.edu/abs/2011ApJ...737...21L},
	author = {{Leung}, Po Kin and {Gammie}, Charles F. and {Noble}, Scott C.},
	doi = {10.1088/0004-637X/737/1/21},
	eid = {21},
	journal = {\apj},
	month = aug,
	number = {1},
	pages = {21},
	title = {{Numerical Calculation of Magnetobremsstrahlung Emission and Absorption Coefficients}},
	volume = {737},
	year = 2011}
\bibliographystyle{aasjournal}

\end{document}